%% file: main.tex
\documentclass[letterpaper,twocolumn,10pt]{article}
\usepackage{usenix-2020-09}
\usepackage{amsfonts}
\usepackage{tikz}
\usepackage{amsmath}
\usepackage[]{hyperref}
\usepackage[dvipsnames]{xcolor}
\usepackage{xspace}
\usepackage{subcaption}

\newcommand{\sysname}{EMA\xspace}
\newcommand{\pisysname}{Pie\xspace}

\begin{document}

\date{}

\title{\sysname: Elastic and Performance Transparent Memory Across GPUs}

\author{
{\rm Yi Xu}\\
UC Berkeley
\and
{\rm Tian Xia}\\
UC Berkeley
\and
{\rm Ion Stoica}\\
UC Berkeley
}

\maketitle

\begin{abstract}
Multi-GPU servers have become the standard building block of modern data centers, providing aggregated capacity through high-bandwidth interconnects. At the same time, workloads such as LLM inference exhibit highly dynamic memory demands, which can cause one GPU to exhaust its local memory while others remain underutilized. This mismatch motivates a model of elastic resource sharing across GPUs.

We present \sysname{}, a memory sharing system that allows GPUs within a server to borrow and reclaim memory from each other, forming an elastic pool of capacity. \sysname{} ensures performance transparency for both borrowers and lenders. For borrowers, prefetching hides remote access costs so that applications experience remote and local memory as indistinguishable in performance. For lenders, borrowed resources remain reclaimable on demand, guaranteeing that performance never falls below that of static partitioning. While our design focuses on memory, the same principle naturally extends to other GPU resources.

Our evaluation shows that \sysname{} improves individual user throughput by up to 52\%, achieves 96\% of the throughput of a system provisioned with $2\times$ capacity, and maintains latency similar to the static local baseline.

\end{abstract}

\pagestyle{plain}

\input{intro}
\input{background}
\input{design}
\input{memory-slice}
\input{multi-gpu}
\input{eval}

\input{related}
\input{conclusion}

\bibliographystyle{plain}
\bibliography{reference}

\input{appendix}

\end{document}

%% file: intro.tex
\section{Introduction}

In recent years, AI and data analytics increasingly rely on GPU acceleration. Large language models (LLMs), video processing, and graph analytics now enable a wide range of applications, from natural language understanding and recommendation to scientific and real-time media analysis.
These workloads are growing beyond what a single accelerator can handle, both in memory footprint and in computational demand.
To meet this demand, modern deployments place multiple GPUs on the same server. Multi-GPU servers provide the resources needed to host massive models and sustain high throughput, while also enabling efficient intra-server communication through high-bandwidth interconnects like NVLink. Cloud providers typically expose these resources as fixed instance types with predefined bundles~\cite{li2019evaluating_gpu_interconnect, nvidia_dgx_a100, nvidia_gb200_nvl72, nvidia_gb300_nvl72, nvidia_nvlink}.

Despite the breadth of available configurations, many workloads exhibit strong time variation in resource demand.
For example, in LLM inference, the KV cache grows with the number and length of concurrent queries, which creates bursts in memory pressure even when average demand is moderate.
Under today's model of static provisioning, a model running on one device can exhaust its local HBM while other GPUs in the same node remain partially idle.
Because instance assignments are fixed after allocation~\cite{jiang2025demystifying, griggs2024m, qiao2021pollux}, capacity that could alleviate pressure is stranded, which leads to underutilization for some users and shortages for others, as well as unnecessary operational cost.

\begin{figure}
    \centering
    \includegraphics[width=0.7\linewidth]{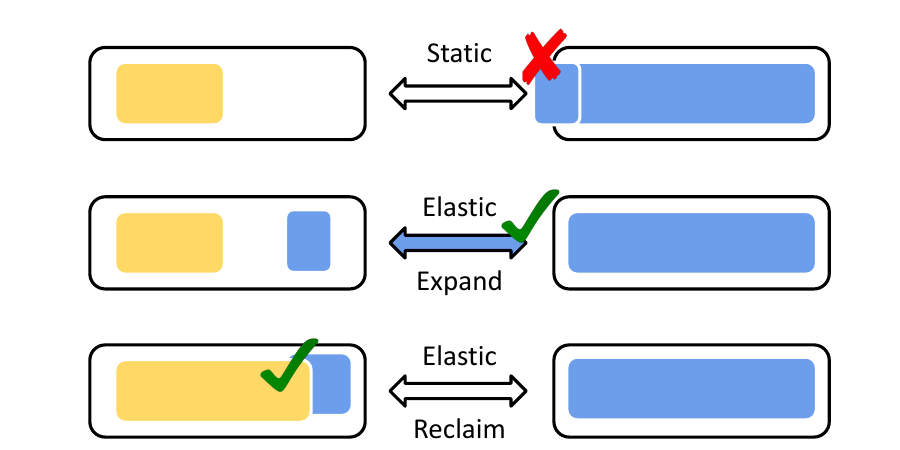}
    \caption{GPU Memory as an Elastic Resource}
    \label{fig:growandreclaim}
\end{figure}

This work pursues a simple idea: when a task runs out of memory on the local GPU, it should be able to borrow the available memory of other GPUs on the same machine. When the lender later needs that memory, it can reclaim it (Figure~\ref{fig:growandreclaim}).
The key is that such borrowing must be performance transparent for both the borrower and the lender. From the perspective of the borrower's application, the borrowed memory is exposed as part of local memory, preserving the application's latency and throughput; we term this property \textbf{Local–Remote Symmetry}. At the same time, the lender must be able to reclaim its full static allocation at any point, so its own application proceeds unaffected, a property we call \textbf{Static–Dynamic Symmetry}. Together, these symmetries enable GPUs to flexibly share capacity without introducing overhead.

We introduce \sysname, a prototype memory system for LLM inference that realizes elastic, performance-transparent memory sharing across GPUs in a node. \sysname organizes memory available to share into \emph{memory slices}.

Our key insight is that a memory slice provides a uniform capacity abstraction across GPUs: for the application, allocations placed in a slice deliver the same performance regardless of whether they reside in local or remote memory.
\sysname forms a memory slice as a dedicated segment of memory shared exclusively between two peers, with each contributing a portion.
Slices are established only across two GPUs with direct links.
With a private link, the available bandwidth is dedicated to the pair.
\sysname sequences transfers across the two peers.
It decides when data moves and how much is sent in each direction.
This sequencing effectively divides the fixed bandwidth between the peers.
As a result, the duration of each transfer becomes predictable.
Leveraging knowledge of the application behavior, the data in the slices is prefetched to the accessing GPU before use, ensuring that remote and local memory accesses are indistinguishable in performance from the application's perspective.

The size of a slice is determined by hardware and application configuration, while the share allocated to each peer is dynamically negotiated at runtime, allowing expansion by borrowing or contraction by reclaiming as needed.

\sysname runs a compute process on each GPU and a memory slice process on each slice. Memory slice processes handle address translation and prefetching.
By aggregating multiple slices, they form an elastic, performance-transparent memory pool that enables each GPU to extend its usable capacity beyond local HBM by borrowing from directly connected peers. On real workloads, \sysname improves throughput by up to 52\% over static partitioning. It achieves 96\% of the throughput of a system provisioned with 2$\times$ KV-cache capacity, while maintaining similar per-token latency.

Although our focus is memory for LLM inference, the approach generalizes to other resources and workloads. Treating a multi-GPU node as a pool that can lend resources on demand points toward cross-GPU virtualization, where the aggregate capacity of a node is flexibly and transparently shared rather than statically partitioned.

In summary, this paper makes the following contributions:
\begin{itemize}

    \item We identify the limitations of existing GPU resource management approaches and introduce new principles for memory sharing in multi-GPU environments.

\item We present \sysname{}, the first system to enable performance-transparent memory sharing across GPUs.

\item We introduce the memory slice abstraction, which ensures that from the borrower's perspective remote memory behaves as local, and from the lender's perspective it is as if the borrowing had never occurred.

    \item We perform a set of evaluations of \sysname on real-world LLM workloads, demonstrating its effectiveness over existing solutions.
\end{itemize}

%% file: background.tex
\section{Background}
\label{sec:background}

Next, we describe the inherent conflict between the fluctuating nature of GPU workloads and the static partitioning methods adopted by cloud providers.
This mismatch leads to missed opportunities to optimize resource utilization and accommodate more workloads within the same hardware configuration.

\subsection{Fluctuation of GPU Workloads}
\label{subsec:gpu-fluctuation}

GPU resource usage fluctuates across tasks due to workload dynamics.
In neural network training, variations in data complexity—such as sequence lengths in NLP or input resolutions in vision, change memory and compute demands even at fixed batch size~\cite{paszke2019pytorch,micikevicius2018mixed}.
Real-time simulations fluctuate with environment detail, physics, and user interactions while meeting strict latency constraints~\cite{mirtich1996impulse,nvidia_physx}.
Video rendering varies with scene complexity (objects, lighting, textures, effects), with adaptive resolution and tile-based rendering adjusting resource allocation~\cite{unrealengine,pharr2016physically}.

In particular, as a widely adopted solution across various applications~\cite{pixelplex2023applications, indatalabs2023applications, ubiops2023applications}, LLM inference also exhibits fluctuations in memory usage. Most of the GPU memory is allocated to store large model weights and the KV cache. While the size of the model weights remains constant, the memory footprint of the KV cache is dynamically allocated, growing and shrinking over time.

\begin{figure}[t]
    \centering
    \includegraphics[width=0.9\linewidth]{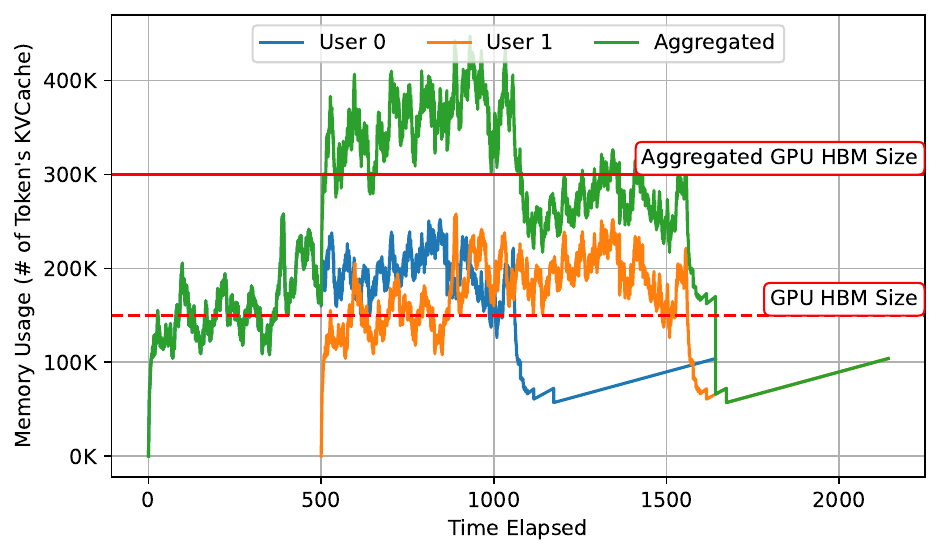}
    \caption{
Memory Utilization Variation.
We simulate memory usage for LLM inference with a maximum batch size of 256 on the ShareGPT workload.
The blue and orange lines show per-user usage, while the green line shows aggregate usage across users.
Aggregate usage exceeds total capacity far less often than static partitioning exceeds individual capacities.
}
    \label{fig:mem_usage_two_sharegpt}
\end{figure}

We show the memory usage fluctuations of two inference engines in Figure~\ref{fig:mem_usage_two_sharegpt}.
The memory usage increases significantly with a higher number of concurrent requests or longer output sequences, as each token's KV Cache consumes memory proportional to the model dimension.
This can quickly exhaust the allocated GPU memory.
In contrast, when a request completes processing, the associated KV cache memory is fully released, resulting in a noticeable drop in memory utilization. The magnitude of this drop varies depending on the total length of the processed request.
These fluctuations in memory allocation and release make the memory utilization of LLM inference highly volatile and unpredictable.

For simplicity, we assume that each user exclusively occupies a single GPU. In scenarios where multiple users statically share a GPU, the same analysis applies by treating their aggregated resource demand as a single composite user.
Throughout the rest of this paper, we use the terms ``user,'' ``model,'' and ``GPU'' interchangeably.

\subsection{Inefficiency of Static Partition}

Modern data centers increasingly deploy multi-GPU nodes, where multiple high-performance GPUs (e.g., A100, H100) share a motherboard and are interconnected via high-bandwidth NVLink~\cite{nvidia_dgx_a100, nvidia_h100, nvidia_a100}. These nodes, common in cloud platforms such as AWS~\cite{aws_cloud}, Azure~\cite{azure_cloud}, and GCP~\cite{GoogleCloudPlatform}, as well as in on-premises systems~\cite{kalra2024comparative, younus2024systematic}, typically house 2 to 8 GPUs per node.
Emerging rack-scale platforms push this idea much further: NVIDIA's Grace–Blackwell NVLink systems (GB200/GB300) wire entire racks into a single NVLink domain. A GB200 NVL72 integrates 36 Grace CPUs with 72 Blackwell GPUs in a rack, while the GB300 NVL72 similarly unifies 36 Grace CPUs with 72 Blackwell \emph{Ultra} GPUs~\cite{nvidia_gb200_nvl72,nvidia_nvlink,nvidia_gb300_nvl72}.

Such nodes enable in compute-intensive tasks like deep learning training, HPC simulations, and large-scale inference. For lighter workloads, users often rent GPU subsets in integral units on the same node~\cite{aws_cloud}, typically with a minimum of one GPU on most cloud platforms. These subsets are statically partitioned for the duration of a user's session. Although partitions may change when users leave and new ones arrive, they remain fixed during extended sharing periods, leading to several drawbacks.

\paragraph{External Fragmentation.}
Although Azure offers fractional GPUs as an option~\cite{azure_gpu_partitioning}, these instance types are restricted to a specific GPU model (A10) and require specialized drivers for proper functionality.
Otherwise, users are required to allocate an integer multiple of GPUs, which reduces flexibility, particularly when the estimated resource requirements do not align with the capacity of a single GPU. For example, hosting models that require only 30 GB of memory (including weights and KV cache for maximum concurrency) on an A100 GPU with 40 GB of device memory result in 10 GB of external fragmentation—unused memory that remains inaccessible to the user.

\paragraph{Internal Fragmentation.}
Even within the memory allocated to the user, the fluctuations discussed in Section~\ref{subsec:gpu-fluctuation} can still lead to internal fragmentation, leaving portions of memory unused and idle at times.
In Figure~\ref{fig:mem_usage_two_sharegpt}, these internal fluctuations, occurring at varying rates, can lead to periods where statically allocated memory experiences shortages for $49.5\%$ of the time, while resources allocated to other users remain underutilized. However, we observed that the aggregated memory usage of the two users is more likely to stay within the total memory capacity for $27.5\%$ of the time, highlighting a potential improvement in resource sharing efficiency.

\paragraph{Wasted Bandwidth.}

The NVLink interconnect within a node typically utilizes mesh topologies, enabling GPU-to-GPU bandwidths of hundreds of GB per second. For instance, NVLink 2.0 on V100 GPUs achieves over 300 GB/s, while A100 and H100 GPUs can exceed 600 GB/s and 900GB/s in certain configurations~\cite{nvidia_dgx_a100, nvidia_h100}. More importantly, these connections exist between every arbitrary pair of GPUs, resulting in $\binom{n}{2}$ dedicated links for a system with $n$ GPUs.
In GB200 NVL72 systems, NVLink Switch exposes all 72 GPUs as one logical accelerator. Each GPU can inject and receive up to about 900~GB/s into the fabric in each direction ($\approx$1.8~TB/s bidirectional), shared across communication with its peers.
However, under static partitioning, the bandwidth between GPUs assigned to different users remains entirely idle, as the isolation abstraction provided by cloud environments ensures no communication across user boundaries. These unused interconnect resources present an opportunity to enable memory sharing across users.

\subsection{Performance Transparent Swapping}
\label{subsec:pie-background}

Accessing memory beyond the physical local HBM capacity poses significant challenges. This is due to the high overhead commonly associated with traditional swapping mechanisms.

\pisysname~\cite{xu2024pie}, as a prototype implementation aimed at addressing this challenge, facilitates memory capacity expansion for LLM inference through performance-transparent swapping.
\pisysname dynamically partitions KV caches into two parts, storing some layers on the GPU's HBM and others on the CPU's DRAM.
Due to the sequential and cyclic access pattern across layers during LLM inference, \pisysname observes that a given layer will not be accessed again until all other layers have been accessed.
\pisysname exploits this \textit{latency window} between two consecutive accesses to the same layer to perform swapping: as long as the data is swapped back before the next access, the application does not perceive that a swap-out occurred. This effectively increases the available memory capacity without incurring performance overhead.

By pipelining data transfer with computation, \pisysname overlaps compute latency with swapping latency, creating the illusion of a larger external memory that behaves with the same performance characteristics as the GPU's native HBM.
The size of this \textit{transparent expanded memory} is determined by the product of compute latency and interconnect bandwidth, which defines the maximum amount of data that can be transferred between two consecutive accesses to the same layer.
The \textit{expansion ratio} is defined as the ratio of total usable memory, consisting of performance-transparent remote memory and the local HBM portion, to the capacity of that local HBM portion.

%% file: design.tex
\section{A Path to Cross-User Memory Sharing}
\label{sec:design}

To improve GPU usage efficiency, we advocate for memory sharing across users.
We first define two objectives, static–dynamic symmetry and local–remote symmetry, which capture the guarantees required for performance-transparent sharing.
We then articulate two principles that make these objectives achievable: applications must expose preemptable points and tolerate some level of information exchange.

Static partitioning simplifies system design and guarantees predictable performance, but it cannot adapt to the dynamic nature of GPU workloads, leaving memory underutilized at times and insufficient at others. Sharing unused capacity across users can smooth out usage spikes and improve efficiency, as illustrated in Figure~\ref{fig:mem_usage_two_sharegpt}. Our goal is to achieve elastic and transparent sharing, which we formalize as a double-symmetry objective.
\label{symmetry-target}

\begin{itemize}
    \item \textbf{Static-Dynamic Symmetry}: Each user must maintain performance that is at least equivalent to what they would achieve under a static allocation.
    \item \textbf{Local-Remote Symmetry}: From a performance perspective, access to remote and local resources should be indistinguishable to users.
\end{itemize}

To achieve these goals, cross-user memory sharing must follow certain principles.

\paragraph{Preemptible Semantics.}

Any resources shared from one user must remain ``preemptible,'' meaning the user can reclaim those resources on demand. This ensures that each user retains their original static allocation as a guaranteed lower bound, even in shared environments. To enable this, applications must define clear preemption points - specific stages in the workload process where interruptions will not cause inconsistency and can safely allow for exit or recovery. The more frequent these preemption points, the smoother the resource sharing process becomes. Most GPU workloads inherently support this: LLM inference can be preempted at any time, training workloads can be interrupted between batches or epochs, video processing can pause between frames, and graph analytics can halt between algorithm iterations.

\paragraph{Information Exchange.}
Applications must tolerate a certain level of information exchange regarding their resource requirements and operational characteristics. The access patterns of the resource recipient can influence the extent to which resources can be shared transparently, and the lending user might observe variability in resource availability as a result. Similarly, applications experiencing different levels of resource shortages may benefit unequally—for instance, a heavily memory-bound process gains significantly from additional memory, whereas a less memory-bound process does not. To enable efficient and fair resource allocation under this principle, applications must share key resource demands and operational characteristics.
\begin{figure}[t]
    \centering
    \includegraphics[width=0.8\linewidth]{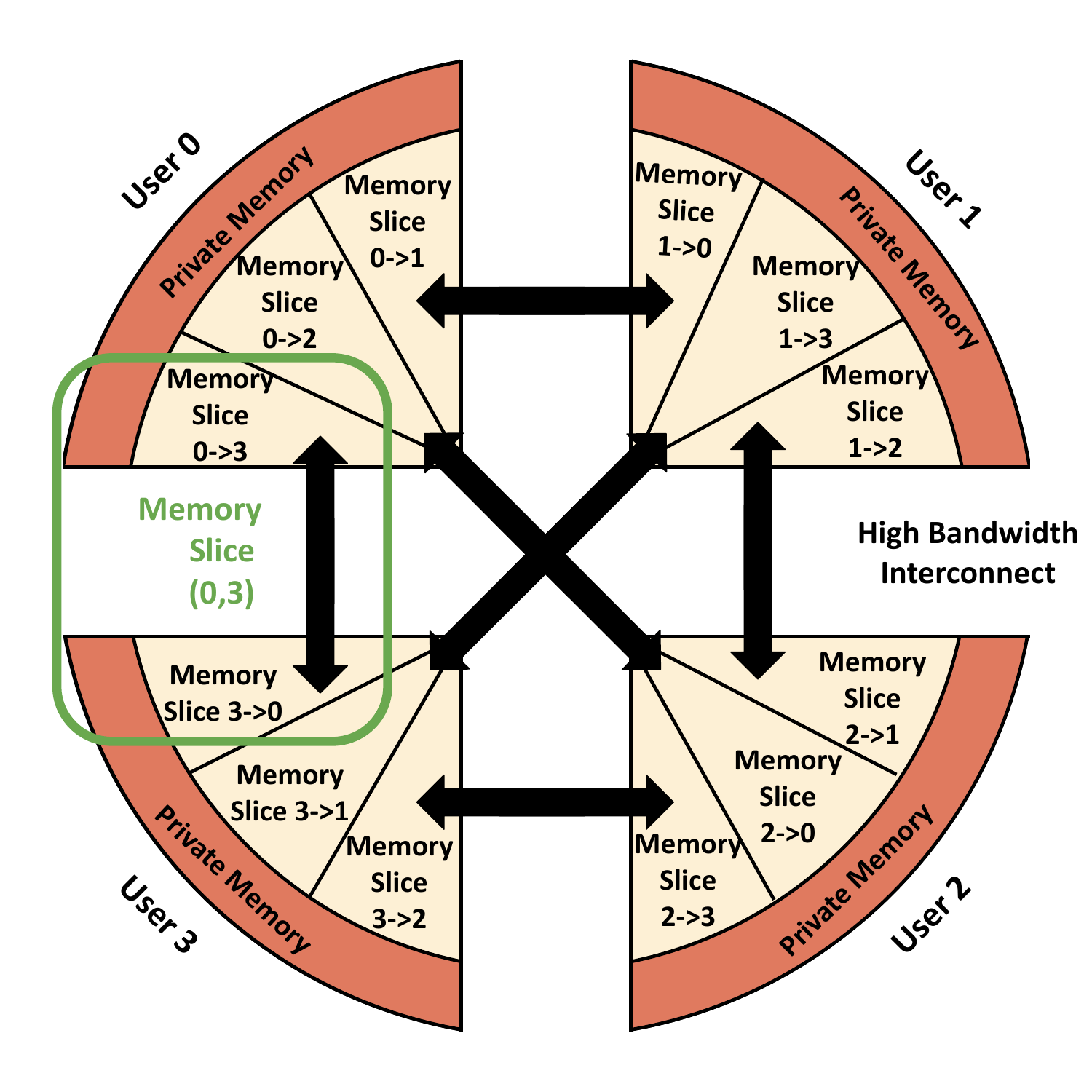}
    \caption{An Illustration of \sysname.}
    \label{fig:evm}
\end{figure}

\sysname leverages the fact that these principles naturally hold in LLM inference to realize the double-symmetry objective. Preemption is always feasible because queries are stateless and idempotent, allowing a user to reclaim previously borrowed memory by preempting requests from others that are using it. For information exchange, the system only needs to expose the latency window and changes in dynamic memory requirements over time. This exchange reveals limited information about workload behavior, and it is used solely for coordination within \sysname{} rather than being broadly exposed across GPUs. Our prototype demonstrates the effectiveness of these principles in practice.

As shown in Figure~\ref{fig:evm}, the user memory space is divided into two parts: private memory and memory slices. We define private memory as the memory that is dedicated to one user, and memory slice segments as chunk of memory each user contributes for sharing.

%% file: memory-slice.tex
\section{Memory Slice: Sharing Between Two GPUs}

The memory slice abstraction serves as the foundation for cross-user memory sharing.
Its structure ensures static–dynamic symmetry and local–remote symmetry: each GPU always retains at least the static portion of memory it contributed to the slice, and the slice delivers uniform performance to applications regardless of whether a given location resides in local or remote memory.

Each memory slice is exclusively shared between two users that are directly connected via a high-speed interconnect such as NVLink, with the associated bandwidth dedicated solely to their use.
Each slice is jointly reserved, granting both users the ability to use it as available and needed.
Individual memory segments within the slice are assigned to a single user on demand and may be reassigned over time.
When a segment is assigned, a virtual-to-physical mapping table ensures that only the assigned user can access it.

Formally, we denote a memory slice using the indices of the GPUs that contribute to it. Suppose GPU i and GPU j are directly interconnected with bandwidth B. We define the shared memory slice as $\mathsf{mem\_slice}(i, j)$, which consists of two parts: $\mathsf{mem\_slice}_{i \rightarrow j}$: the portion contributed by GPU i's HBM, and $\mathsf{mem\_slice}_{j \rightarrow i}$: the portion contributed by GPU j's HBM.
For example, $\mathsf{mem\_slice}_{0 \rightarrow 3}$ and $\mathsf{mem\_slice}_{3 \rightarrow 0}$ together form $\mathsf{mem\_slice}(0,3)$, as shown in Figure~\ref{fig:evm}.

\begin{figure}[t]
    \centering
    \includegraphics[width=0.8\linewidth]{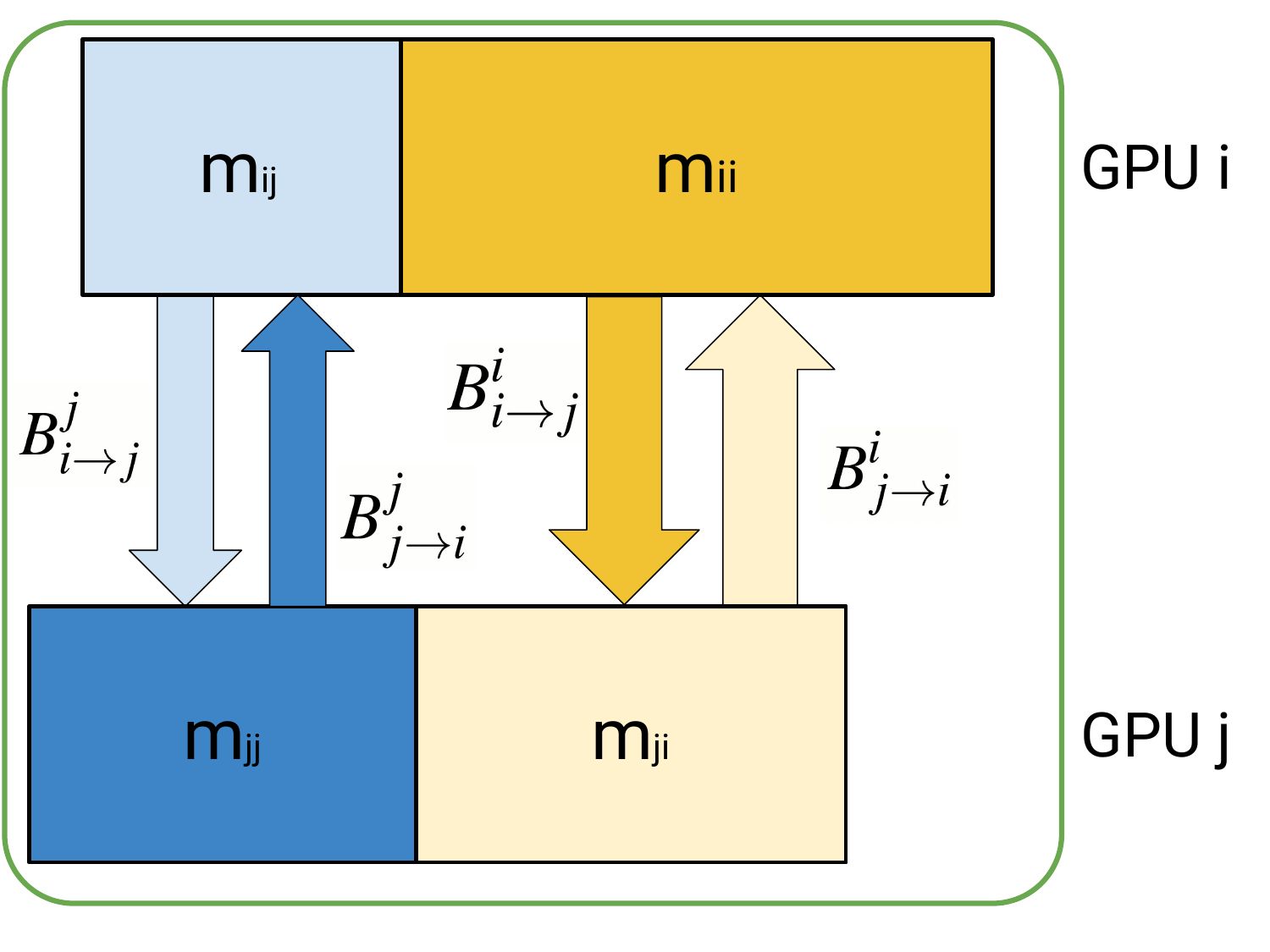}
    \caption{Memory and Bandwidth Sharing in a Memory Slice. The upper half shows physical GPU~$i$ and the lower half shows physical GPU~$j$, which are directly interconnected with a certain bandwidth. The yellow region belongs logically to GPU~$i$, and the blue region belongs logically to GPU~$j$.
    }
    \label{fig:bw-mem-sharing}
\end{figure}

Regardless of where the data physically reside, both GPU i and GPU j can store the data in $\mathsf{mem\_slice}(i, j)$ and access them with performance comparable to accessing their respective local HBM.
Figure~\ref{fig:bw-mem-sharing} shows an example of memory slice sharing between GPU $i$ and GPU $j$.

Assume that $\mathsf{mem\_slice}_{i \rightarrow j} = m_{ii} + m_{ij}$, where $m_{ii}$ is the portion of memory actually utilized by GPU $i$, and $m_{ij}$ is the portion utilized by GPU $j$.
Similarly, let $\mathsf{mem\_slice}_{j \rightarrow i} = m_{ji} + m_{jj}$, where $m_{ji}$ is the portion utilized by GPU $i$, and $m_{jj}$ is the portion utilized by GPU $j$.

The unidirectional bandwidth from GPU $i$ to GPU $j$ is $B_{i \rightarrow j}$, which is composed of two components:
$B_{i \rightarrow j} = B_{i \rightarrow j}^i + B_{i \rightarrow j}^j$
where $B_{i \rightarrow j}^i$ and $B_{i \rightarrow j}^j$ represent the bandwidth portion that GPU $i$ and GPU $j$ can utilize, respectively.
Similarly, the unidirectional bandwidth from GPU $j$ to GPU $i$ is:
$B_{j \rightarrow i} = B_{j \rightarrow i}^i + B_{j \rightarrow i}^j$
with $B_{j \rightarrow i}^i$ and $B_{j \rightarrow i}^j$ defined in the same way.

Assuming GPU $i$ have a latency window $L_i$ (see Section~\ref{subsec:pie-background}), and GPU $j$ have a latency window $L_j$.
To ensure that swapping remains transparent, the data transfer must complete within the available latency window. This requires the following conditions to hold:
\begin{align*}
B_{j \rightarrow i}^i \times L_i \geq m_{ji} \\
B_{i \rightarrow j}^j \times L_j \geq m_{ij}
\end{align*}

In order to swap in a certain amount of data, the local GPU must have sufficient memory space to accommodate incoming data. This implies that, in order to fully utilize available bandwidth, during the same time interval, the amount of data swapped in must equal the amount of data swapped out.
As a result, the following constraints must also hold:
\begin{align*}
B_{i \rightarrow j}^i \times L_i \geq m_{ji} \\
B_{j \rightarrow i}^j \times L_j \geq m_{ij}
\end{align*}

Given the bandwidth decomposition:
\[
B_{j \rightarrow i}^i + B_{j \rightarrow i}^j = B_{j \rightarrow i}, \quad
B_{i \rightarrow j}^i + B_{i \rightarrow j}^j = B_{i \rightarrow j}
\]

We arrive at the following conclusion (full derivation provided in Appendix~\ref{sec:proof}):

\begin{equation}
\frac{m_{ji}}{L_i} + \frac{m_{ij}}{L_j} = B_{j \rightarrow i} = B_{i \rightarrow j} = B
\label{eq:bandwidth-balance}
\end{equation}

The terms \( m_{ji} \) and \( m_{ij} \) represent the physical memory capacity that one GPU has borrowed from the other.
If \( m_{ji} = m_{ij} \), then the borrowing is effectively neutralized, as each GPU lends and borrows the same amount, making the sharing redundant.
Meaningful sharing occurs only when there is a significant difference between \( m_{ji} \) and \( m_{ij} \).
This raises the question of how much memory each GPU should borrow from the other.
The decision must be guided by the double symmetry objectives: static-dynamic symmetry and local-remote symmetry.

\subsection{Static-Dynamic Symmetry}
At any given time, each GPU is guaranteed having access to at least its original physical memory capacity, regardless of whether the allocated memory resides locally or remotely.
For LLM inference, workloads can be preempted anytime, allowing borrowing and reclamation to happen immediately as needed.
Reclamation operates on logical ownership: a GPU regains capacity by preempting regions that are logically owned by its peer, irrespective of whether those regions reside on its own device or on the peer.

If GPU \(j\) preempts GPU \(i\) to reclaim \(k\)~GB, then GPU \(j\) regains \(k\)~GB from \(m_{ji}\) or \(m_{ii}\). In theory, this allows \(m_{jj}\) or \(m_{ij}\) to increase by \(k\).
In practice, data at different locations may be dependent with each other. Suppose GPU~$i$ has an expansion ratio $r$, which means that its effective memory capacity is $r$ times its local HBM.
In the context of LLM inference, making memory expansion performance-transparent corresponds to each request placing $m$ layers remotely and $n$ layers locally. The ratio $(m+n)/n$ must match the expansion factor $r$ (see Section~\ref{subsec:pie-background}).
When GPU~$i$ releases $k$~GB from the remote side, there remain $(r-1)\cdot k$~GB of local data from the same request but different layers. These local entries are only useful if the corresponding remote KV cache is present; otherwise, retaining them wastes capacity. We therefore treat the $(r-1)\cdot k$~GB as being in the dependency chain of the reclaimed $k$~GB, and require GPU~$i$ to release it as well.

If \((r - 1) \cdot k > k\), the system releases more local memory than can be reborrowed remotely, resulting in \textit{wasted capacity} that could have otherwise been used by GPU \(j\)'s data. This excess memory must either be garbage collected into GPU \(i\)'s private memory or left underutilized.

On the other hand, if \((r - 1) \cdot k < k\), then GPU \(i\) lacks sufficient local memory to support the full \(k\)~GB increase in \(m_{ij}\), leading to \textit{underutilized bandwidth} because there is not enough memory capacity for data to fully saturate the transfer.

Therefore, to avoid frequent garbage collection and to keep both memory and bandwidth fully utilized, we enforce the condition \((r - 1) \cdot k = k\), which yields a fixed expansion ratio of \(r = 2\). Because the expansion ratio is fixed at 2, we know that \(m_{ij} = m_{jj}\) and \(m_{ii} = m_{ji}\), which leads to \(m_{ij} + m_{ii} = m_{jj} + m_{ji}\), meaning that \textbf{both GPUs contribute an equal amount of memory to the memory slice}.

Memory slice sizes are negotiated collaboratively, and the allocation ratio can adapt dynamically throughout the sharing period.
Given an expansion ratio of 2 for both users, when GPU \(j\) reclaims \(k\)~GB from GPU \(i\), GPU \(i\) releases \(k/2\)~GB on GPU \(j\), and simultaneously releases \(k/2\)~GB on its own local memory.
This is because releasing \(k/2\)~GB on GPU \(j\) corresponds to freeing half of the associated layers, making the corresponding \(k/2\)~GB on GPU \(i\) no longer useful and therefore also released.
This implies that, for a memory slice of total capacity \(\mathsf{mem\_slice}(i, j)\), each GPU adjusts its memory usage by modifying its share within this unified address space.
Building on Eq.~\eqref{eq:bandwidth-balance}, we generalize it to capture changing shares, where $m_v$ can take any value in the range $0 \leq m_v \leq m_{ji} + m_{ij}$:

\[
\frac{m_{ji}-m_v}{L_i} + \frac{m_{ij}+m_v}{L_j} \leq B .
\]
\[
\frac{m_{ji}+m_v}{L_i} + \frac{m_{ij}-m_v}{L_j} \leq B .
\]

We conclude that setting $L = \min(L_i, L_j)$ yields the maximum capacity of a memory slice, given by (see detailed proof in Appendix~\ref{proof:per-gpu-contribution}):

\[
\mathsf{mem\_slice}(i, j) = 2 \times (m_{ji} + m_{ij}) = 2 B \cdot L
\]

With this memory layout, satisfying the static--dynamic symmetry is straightforward. When a GPU requires additional memory, it first attempts to allocate from free memory. If the allocation succeeds, it proceeds with using the memory. If the allocation fails, the GPU checks whether it is currently using at least as much memory as its statically allocated capacity.
If not, indicating that it previously lent underutilized memory to the other GPU (i.e., for GPU \(i\), \(m_{ji} < m_{ij}\)), it can reclaim the borrowed portion by preempting a number of requests on GPU \(j\). This is done by asking the memory allocator to identify and evict appropriate requests of GPU \(j\) until GPU \(i\) has reclaimed enough memory to satisfy the allocation request, or the reclamation has made its usage achieve its static capacity (i.e., \(m_{ji} = m_{ij}\)).

\subsection{Local-Remote Symmetry}

Memory slices are designed for dynamic sharing between two users, based on their memory demands over time.
As workloads progress, memory is allocated and deallocated as needed.
This dynamic behavior leads to memory layouts where segments belonging to different users become interleaved within each slice.
However, this interleaving poses a challenge. Previously, we derived the condition
\[
{m_{ji}} + m_{ij} = B \cdot L
\]
By static-dynamic symmetry, we have $m_{ji} = m_{ii}$, and substituting gives
\[
{m_{ii}} + m_{ij} = B \cdot L
\]
This shows that the capacity each GPU contributes to the slice (e.g., for GPU~$i$, $m_{ii} + m_{ij}$) is determined jointly by the effective bandwidth $B$ and the latency windows $L_i$ and $L_j$ of both users.

When data from different GPUs are interleaved within a memory slice, each swap can retrieve only a small fragment at a time.
These small transfers often lead to lower effective bandwidth $B$, as memory copy overhead dominates.
To preserve transparency in such cases, the system must either reduce the usage of remote memory for each GPU or improve interconnect bandwidth efficiency.
To maintain high interconnect bandwidth efficiency, data copy operations must be performed in large chunks.
\sysname{} applies a slab allocator to keep each user's data physically contiguous. It then exploits application's predictable access patterns to group temporally adjacent data (e.g., consecutive layers) together. This way, even when data from different users are interleaved, they can still be transferred in relative large chunks with high bandwidth utilization.

\subsubsection{Layer-major Allocation}
\sysname uses a slab allocator when managing memory slice. Slabs are relatively large, contiguous memory chunks composed of tens to hundreds of KV Cache blocks. Similar to blocks, slab allocation is applied uniformly across all layers.
Each slab is allocated exclusively to a user. When a task exhausts its allocated memory, it requests additional slabs from the slice. In contrast, when a task releases memory upon completing its operations, the freed slabs are returned to the slice for reallocation to either user.
When performing reclamation to enforce static--dynamic symmetry, the memory allocator simply selects a slab belonging to the borrowing GPU that contains the fewest number of active requests.

\begin{figure*}[t]
    \centering
    \begin{subfigure}{0.32\textwidth}
        \centering
        \includegraphics[width=\linewidth]{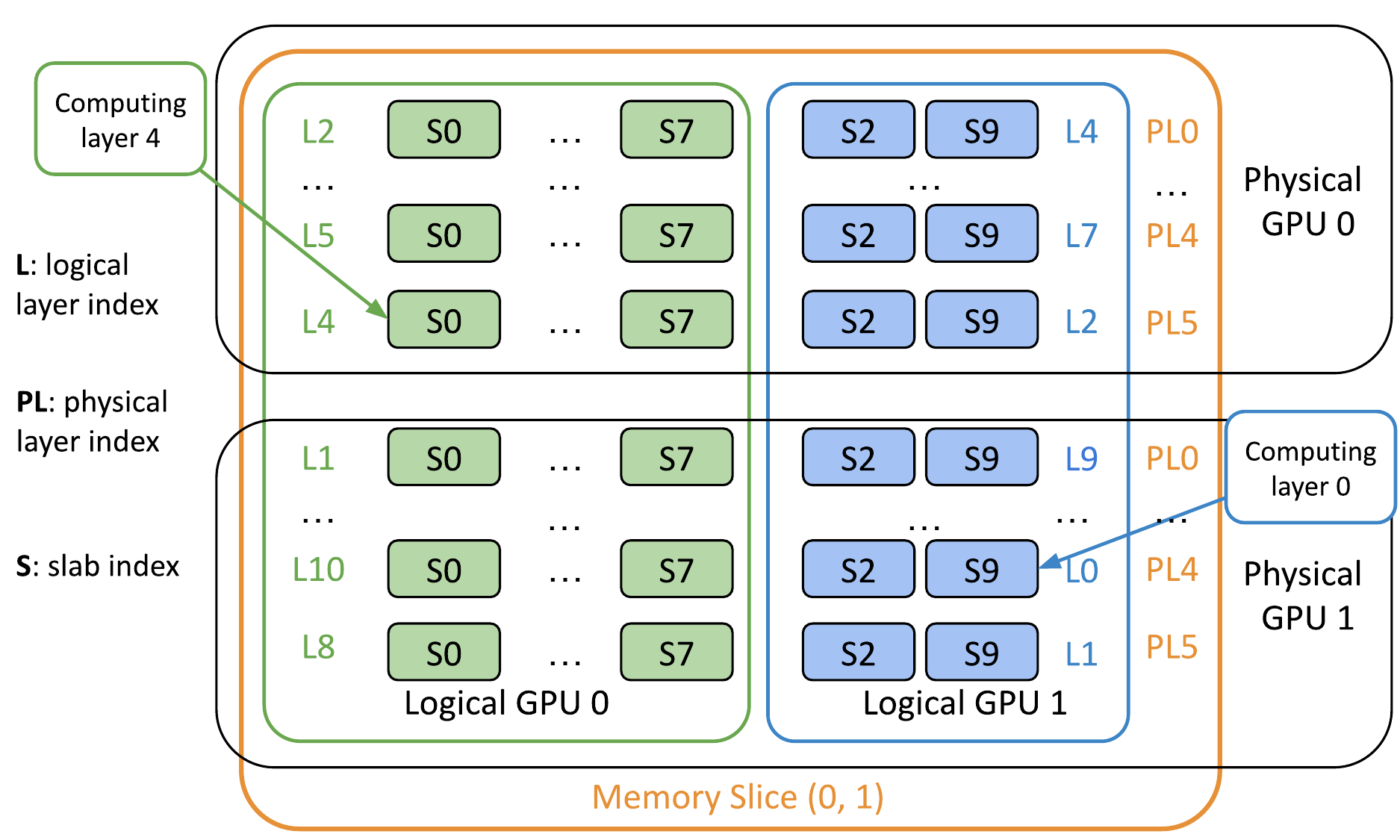}
        \caption{Layer-major Allocation}
        \label{fig:mem-slice-a}
    \end{subfigure}
    \hfill
    \begin{subfigure}{0.32\textwidth}
        \centering
        \includegraphics[width=\linewidth]{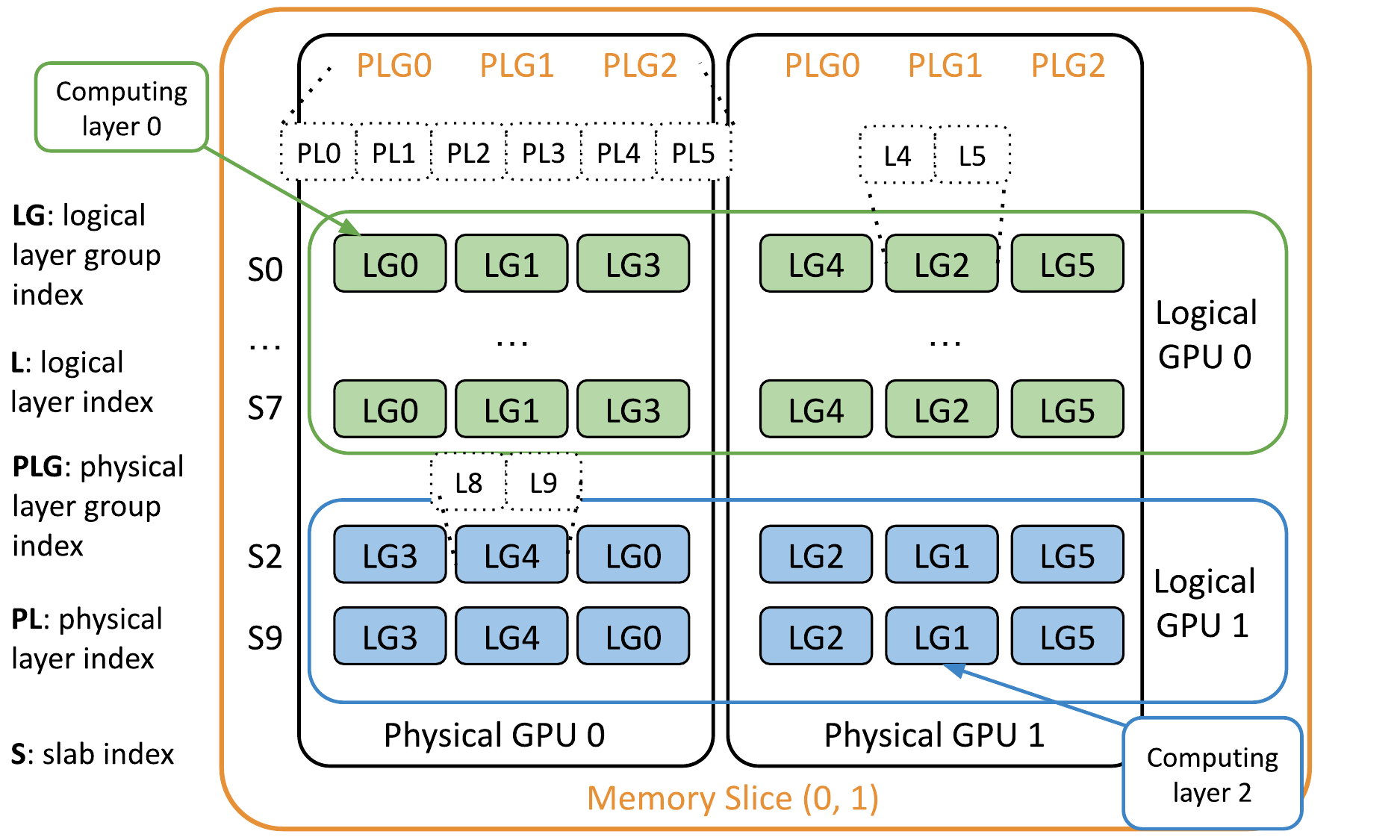}
        \caption{Slab-major Allocation}
        \label{fig:mem-slice-b}
    \end{subfigure}
    \hfill
    \begin{subfigure}{0.32\textwidth}
        \centering
        \includegraphics[width=\linewidth]{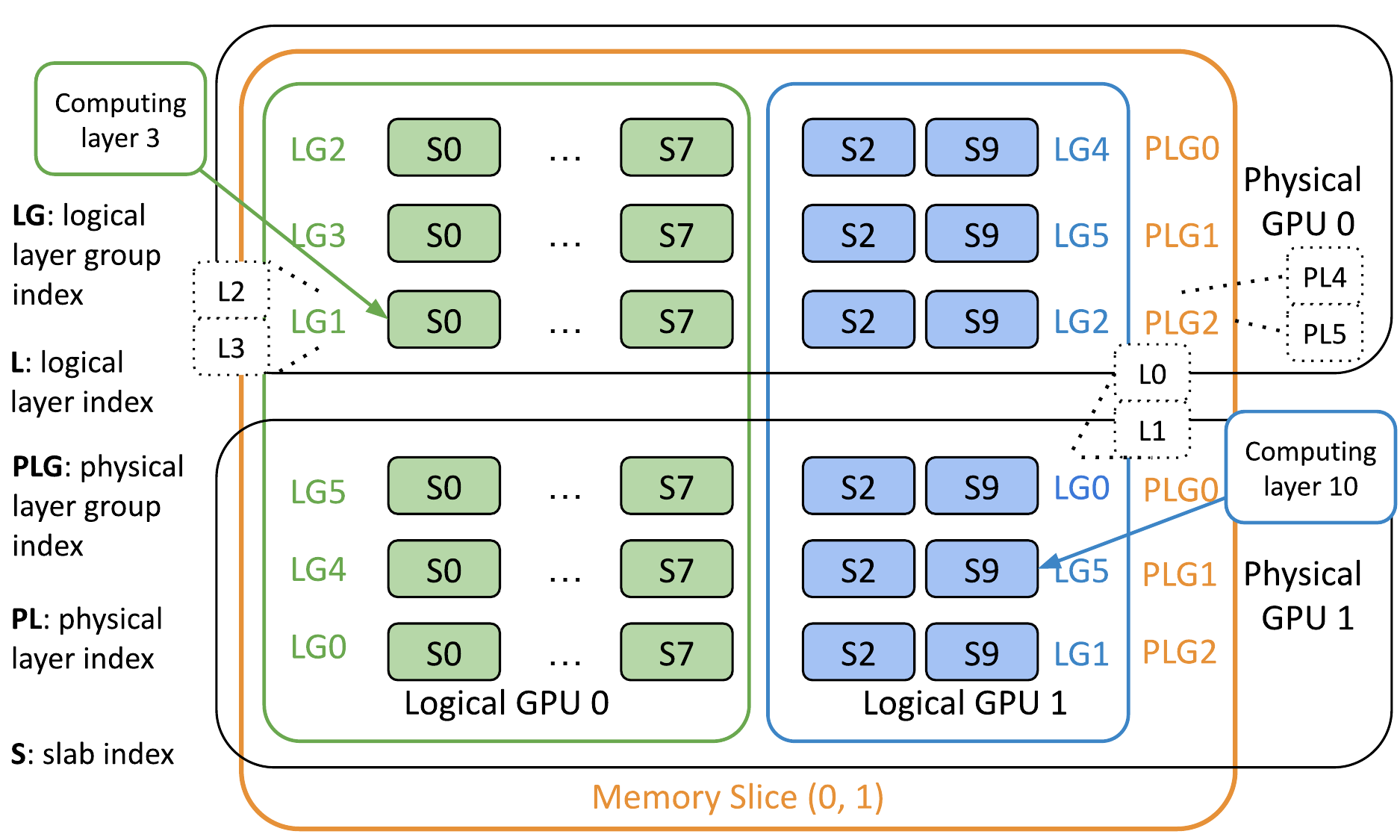}
        \caption{Layer-group-major Allocation}
        \label{fig:mem-slice-c}
    \end{subfigure}
    \caption{Memory Layout in Memory Slice.}
    \label{fig:mem-slice}
\end{figure*}

As shown in Figure~\ref{fig:mem-slice-a}, the memory slice (0,1) spans across the physical HBM of both GPU 0 and GPU 1.
When computation begins,
during allocation, each GPU gets allocated a number of slabs, with half of the layers residing on GPU 0 and others on GPU 1.
During computation, each GPU ensures that the slabs of the upcoming layers are prefetched in time.

This approach enables memory sharing between two GPUs while keeping each GPU's data relatively contiguous, but it must balance two opposing constraints: when workloads are highly dynamic, slab contiguity decreases, and in the worst case neighboring slabs always map to the other GPU, forcing swaps at single-slab granularity; because swapping now occurs one slab at a time, slabs must be large enough to fully utilize interconnect bandwidth. On the other hand, slabs cannot be too large or they suffer from internal fragmentation, transfer unused data during swapping, and reduce sharing efficiency. Since slabs are private to each GPU, any unused space cannot be allocated to the other GPU until the entire slab becomes free and reallocated as a whole, so if slabs are too large, sharing is rarely triggered.

\subsubsection{Slab-major Allocation}
To address this challenge,
\sysname introduces the concept of a \textit{layer group}, which refers to a set of layers that are logically consecutive. \sysname ensures that these layers are also physically contiguous in memory.
As a result, from both the logical and physical perspectives, a layer group behaves like a single large layer for the purpose of swap management.

Current LLM inference engines do not ensure physical contiguity across layers.
As shown in Figure~\ref{fig:mem-slice-a}, memory is contiguous only within a layer, where slab $i$ and slab $i{+}1$ are consecutive, while logically consecutive layers are often not.
Since intra-layer contiguity is a redundant property of layer-major allocation, \sysname{} adopts a slab-major strategy to make logically consecutive layers physically contiguous by placing all layers of a slab together.

We maintain the same logical-to-physical layer mapping, meaning that at any point in time, half of the layers are physically located on the local GPU and the other half on the remote GPU.
To achieve this, we allocate the first half of the layers in a slab contiguously on the local GPU, and the second half contiguously on the remote GPU.

As a result, when a group of layers is swapped, the layers involved are guaranteed to be physically contiguous within each slab, enabling efficient bulk transfers.
Under this strategy, the number of swaps issued for each model is determined deterministically by the number of slabs allocated to it.

With the simplified abstraction of treating each layer group as a ``big layer,'' \sysname applies the same swapping logic as before. The key difference is that, instead of using actual physical and logical layer indices, \sysname operates on layer group indices.
For example, as shown in Figure~\ref{fig:mem-slice-b}, under a layer group size of 2, physical layer group 0--2 corresponds to physical layers 0--5, while logical layer group 2 corresponds to logical layers 4 and 5.

At each layer group boundary, \sysname checks whether swapping should be triggered. Since each swap is performed at the granularity of a layer group, and each scheduling step occurs after the computation of an entire layer group, the latency ratio between swapping and computation remains unchanged. As a result, the performance transparency of the original design is preserved.

This design has two sources of overhead. First, each swap requires a buffer equal to the transferred chunk, which was one layer per model but becomes one layer group with grouping. Second, the effective latency window of LLM inference is proportional to \((k-1)/k\), where \(k\) is the number of layers~\cite{xu2024pie}; with larger groups, \(k\) can drop to 10 or fewer, making the window slightly smaller. Because the expansion size is the product of the latency window and the bandwidth, this reduction directly limits performance-transparent expansion. Thus, while larger groups improve transfer efficiency, they also increase overhead, so group size must be carefully bounded.

\subsubsection{Layer-group-major Allocation}

Slab-major allocation provides the guarantee that swapping occurs at the granularity of a single layer's single slab size multiplied by the layer group size.
However, it enforces contiguity in an unnecessary dimension: it places each layer group contiguously next to the next one.
This is inefficient, as two layer groups are never swapped together.

\sysname transforms the memory layout to take advantage of contiguity in a best-effort manner. Specifically, for slab \(i\) and layer group \(j\), instead of placing slab \(i\), layer group \(j+1\) adjacent to it, we place slab \(i+1\), layer group \(j\) next to it.
We refer to this allocation strategy as \textit{layer-group-major}.
For example, as shown in Figure~\ref{fig:mem-slice-c}, for GPU 0, slab 0 of logical layers 2 and 3 is placed contiguously with slab 1 of those same layers.
This layout enables the swap of slab 0 and slab 1 to be combined into a single transfer.

Under this hierarchical layout, memory contiguity is exploited in two dimensions.
Within each layer group, swapping occurs at least at the granularity of a full group.
Across the slice, \sysname allocates slabs for each model as contiguously as possible.
If one slab holds a full group of size \(kS\) (\(k\): group size, \(S\): per-layer slab size) and each layer spans \(m\) slabs, the effective swap size ranges from \(kS\) to \(mkS\).
This provides a significant improvement over layer-major allocation, where the swapping size ranges from \(S\) to \(mS\), and slab-major allocation, where the swapping size is fixed at \(kS\).
Because transfers occur in large contiguous chunks, this memory layout is able to drive the interconnect close to full utilization in practice.

%% file: multi-gpu.tex
\section{Multi-GPU: Aggregating Memory Slices}

The memory slice serves as the basic unit of sharing. Multiple memory slices can coexist independently on a single GPU, since each GPU typically has multiple independent interconnects with different peers.
As a result, in multi-GPU sharing scenarios, each memory slice operates independently and does not interfere with others.

\subsection{Process Address Space}

\sysname assumes one compute process per GPU, responsible for reading from local HBM and executing computations on model weights, KV cache, and ephemeral data. If a GPU is directly interconnected with $n$ peers willing to participate in memory sharing, it spawns $n$ memory-slice processes, one for each peer.

Each compute process is provided with an \textit{Elastic Address Space (EAS)} abstraction that extends beyond its guaranteed \textit{Local Address Space (LAS)}. The LAS corresponds to the fixed capacity of GPU-local memory, while the EAS includes reclaimable slices contributed by peers. The availability of EAS memory depends on system load and sharing state, so its capacity is not guaranteed. This abstraction allows processes to allocate working sets larger than their LAS while maintaining the ability to fall back on guaranteed local memory.

\begin{figure}[t]
    \centering
    \includegraphics[width=0.9\linewidth]{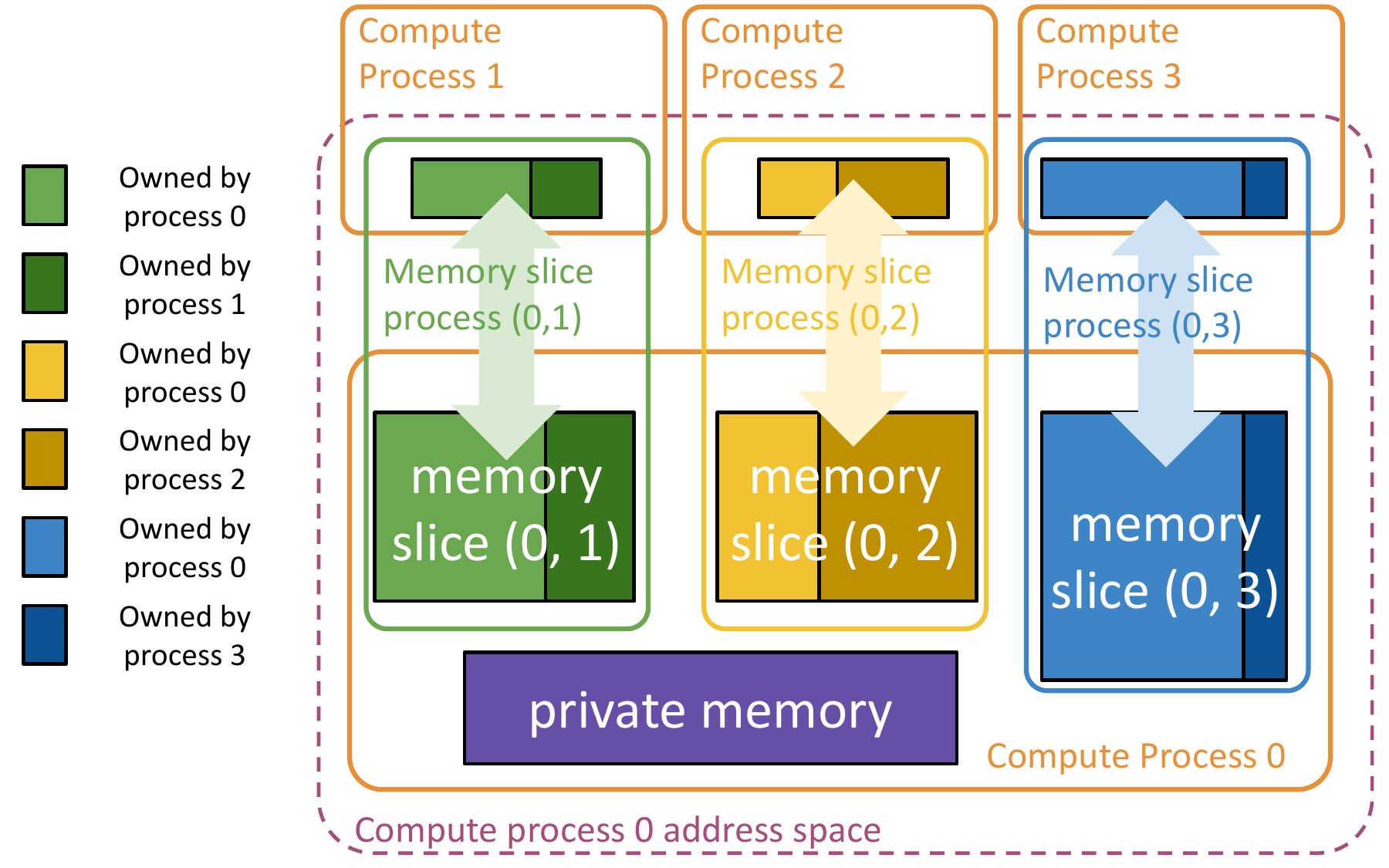}
    \caption{Multiple Memory Slice in Multi-GPU Sharing}
    \label{fig:multi-process}
\end{figure}

Each memory-slice process manages its assigned slice, consisting of equal portions from the two participating GPUs, and exposes it as a shared region accessible to both compute processes. Ownership within the slice is logically partitioned: for example, in Figure~\ref{fig:multi-process}, data in light green are owned by process~0 but physically distributed across GPU~0 and GPU~1, while data in dark green are owned by process~1 with a similarly distributed layout.

A memory-slice process also handles address translation on behalf of its compute process. When an allocation falls within its range, the slice process issues a virtual address, predicts upcoming accesses, allocates a physical location, prepares a directly accessible address, and prefetches the data ahead of time. Thus, when the compute process accesses the memory, it uses the prepared address without additional latency. For objects spanning multiple slices, access proceeds once all relevant slice processes complete preparation; due to the performance-transparent design, this coordination occurs in advance and does not add to the access latency.

\subsection{Scheduling}

A request can arrive at any time and may be scheduled to use any memory slice. Scheduling is not independent: assigning one request to a slice changes how much memory remains available for other processes. For example, if a request on process~0 is scheduled to slice~(0,1), then process~1 will have less memory available, while processes~2 and~3 will have more. Intuitively, scheduling to slice~(0,1) is preferable if process~1 is unlikely to use that memory and processes~2 and~3 are expected to need it.

In \sysname, the key is to avoid driving any slice to full utilization. When a slice reaches 100\%, preemptions occur for two reasons: (1) if the scheduling process already occupies more than 50\%, its own ongoing requests are preempted because capacity is exhausted; (2) if it occupies less than 50\%, static--dynamic symmetry forces reclamation, preempting the peer's requests. Thus, good scheduling should minimize the likelihood that $\text{util}_{k,i} + \text{util}_{k,j} = 100\%$.

Formally, suppose process~$i$ has $n$ slices. For each slice~$k$, let $\text{util}_{k,i}$ and $\text{util}_{k,j}$ denote the utilizations of process~$i$ and its peer~$j$. By construction, $\text{util}_{k,i} + \text{util}_{k,j} \leq 100\%$, and preemption begins once this sum reaches 100\%. We formulate scheduling as an edge-constrained, unknown-length load-balancing problem. To avoid preemption, we adopt a local greedy algorithm that assigns each request to the slice~$k$ with the smallest $\text{util}_{k,i} + \text{util}_{k,j}$. As shown in Appendix~\ref{sec:proof-scheduling}, this algorithm achieves provably good performance bounds.

\subsection{Bursty Workload Preemption Control}

The static–dynamic symmetry can cause excessive preemptions. For example, if user A temporarily exceeds its static capacity, user B may reclaim memory, only for A to use it again when B releases, and then be preempted once more as B's demand rises. This back-and-forth can repeat endlessly.

To prevent such cycles, \sysname caps allocations after a preemption: if a user is preempted when its usage exceeds 50\%, the slice limits that user's future allocations to the preempted level for a sliding window of length $T$. If another preemption occurs within that window at a lower level $k_1 < k_0$, the cap resets to $k_1\%$; if no further preemptions occur, the cap is lifted once $T$ expires. This mechanism reduces disruption under dynamic workloads and high memory pressure, while still allowing aggressive sharing when preemptions are unlikely.

%% file: eval.tex
\section{Evaluation}
\label{sec:eval}
In this section, we answer the following questions:

\begin{figure*}[ht]
    \centering
    \begin{subfigure}[b]{0.24\textwidth}
        \includegraphics[width=\linewidth]{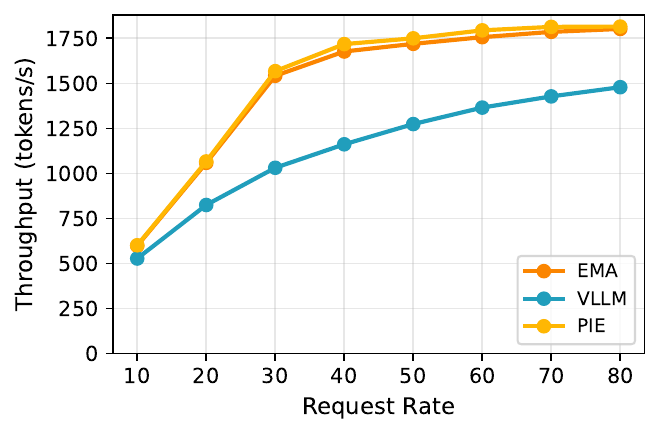}
        \captionsetup{justification=centering}
\caption{LLaMA13B - Alpaca \\ on 4$\times$A100-40GB}
        \label{subfig:llama-alpaca}
    \end{subfigure}
    \hfill
    \begin{subfigure}[b]{0.24\textwidth}
        \includegraphics[width=\linewidth]{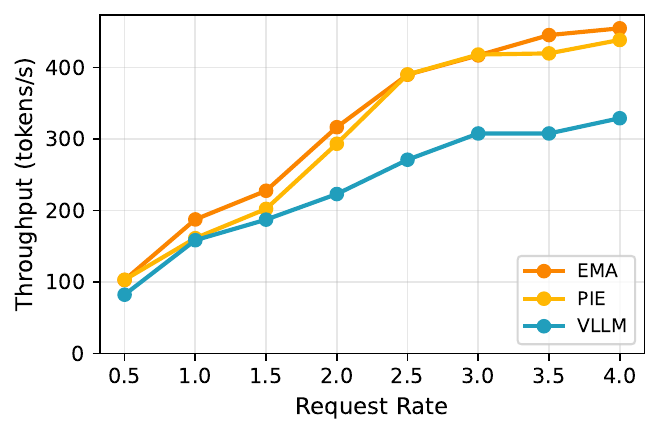}
        \captionsetup{justification=centering}
\caption{LLaMA13B - ShareGPT \\ on 4$\times$A100-40GB}
        \label{subfig:llama13b-sharegpt}
    \end{subfigure}
    \hfill
    \begin{subfigure}[b]{0.24\textwidth}
        \includegraphics[width=\linewidth]{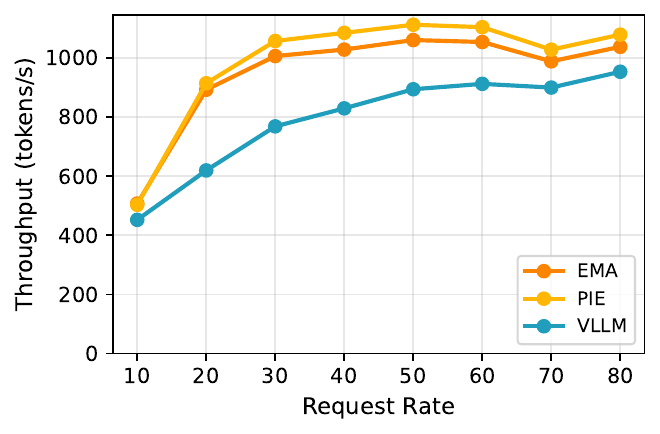}
    \captionsetup{justification=centering}
\caption{OPT30B - Alpaca \\ on 8$\times$A100-80GB}
        \label{subfig:opt30b-alpaca}
    \end{subfigure}
    \hfill
    \begin{subfigure}[b]{0.24\textwidth}
        \includegraphics[width=\linewidth]{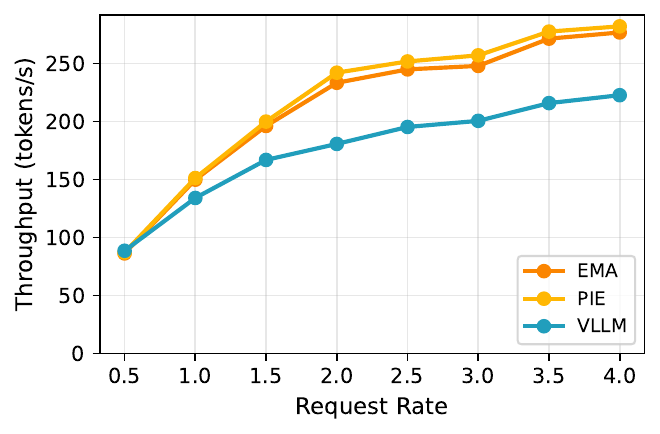}
\captionsetup{justification=centering}
\caption{OPT30B - ShareGPT \\ on 8$\times$A100-80GB}
        \label{subfig:opt30b-sharegpt}
    \end{subfigure}
 
    \caption{Throughput vs. Request Rate on Heavily-Loaded GPUs} 
    \label{fig:workload}
\end{figure*}

\begin{enumerate}
    \item \textbf{Throughput Gains:} How much throughput improvement can \sysname provide compared to a state-of-the-art inference system with statically partitioned resources under different load? (\S\ref{subsec:eval-tput})

    \item \textbf{Memory Utilization:} How much does \sysname improve memory utilization, and how do these improvements affect overall system throughput? (\S\ref{subsec:eval-mem-util})

    \item \textbf{Capacity Expansion and Latency Comparison:} How much performance transparent memory is \sysname able to provide? How does the latency of \sysname compare to the static baseline? (\S\ref{subsec:latency})

    \item \textbf{Local--Remote Symmetry:} How important is local-remote symmetry for memory sharing, and how much performance improvement does this property provide? (\S\ref{subsec:eval-lr})

    \item \textbf{Static--Dynamic Symmetry:} How important is static-dynamic symmetry for memory sharing? How much improvement does it offer globally and for each GPU? (\S\ref{subsec:eval-sd})

   \item \textbf{Memory Slice Study:} How to achieve high performance with the same slice size? (\S\ref{subsec:eval-slice-setup})

\end{enumerate}

\subsection{Experiment Setup}

\paragraph{Hardware Configuration.} 
Each user is assigned a single GPU operating independently. 
The experiments are carried out on NVIDIA servers with 4$\times$A100-40GB and 8$\times$A100-80GB GPUs, where the suffix denotes the local HBM capacity. 
Within each node, all GPU pairs are interconnected via NVLink, offering up to 600\,GB/s bidirectional bandwidth (300\,GB/s per direction under concurrent traffic).

\paragraph{Experiment Setup.} 
Each GPU runs one model independently. 
We evaluate ShareGPT~\cite{sharegpt} and Alpaca~\cite{taori2023stanfordalpaca} on LLaMA~\cite{touvron2023llama} and OPT~\cite{zhang2022opt}, both consisting of real LLM service texts. 
ShareGPT includes longer input and output sequences than Alpaca. 
Request arrivals follow a Poisson distribution with varying rates, and all datasets are tokenized as in prior work~\cite{kwon2023efficient,xu2024pie}.

We compare the performance of the following systems:

\begin{itemize}
\item \textbf{\sysname}: 
We build \sysname{} on top of vLLM~\cite{kwon2023efficient}.
Unless otherwise specified, we allow each GPU to contribute its full available KV cache capacity to its memory slices.
For example, on an A100-40GB running Llama-13B, the model weights occupy approximately 26GB of HBM, and we reserve 1GB for ephemeral data usage, leaving up to 13GB for KV cache.
In a 4-GPU setup, each GPU has 3 peers and can divide its 13~GB contribution equally across them, i.e., $13 \,\text{GB} \div 3 \approx 4.3 \,\text{GB}$ per slice, resulting in a potential expanded capacity of 13~GB for each GPU.
For all experiments except those in Section~\ref{subsec:latency} on latency–expansion, we maintain HBM utilization at 90\%.

\item \textbf{vLLM}: Each GPU uses its full memory and independently hosts a vLLM instance.

\item \textbf{\pisysname{}}: The system was originally designed to offload the KV cache to the CPU. We modified it so that one GPU runs the compute process, while the other GPUs on the same node remain idle and serve purely as passive memory devices.
This configuration represents an upper bound of \sysname{}, as the compute GPU is guaranteed access to 2$\times$ its local capacity for the KV cache.

\end{itemize}

\subsection{Throughput Gains}
\label{subsec:eval-tput}

In this section, we discuss how much throughput improvement can \sysname{} provide compared to the static version backed by vLLM on each GPU.

We fix the request rates for models 0–6 at 10 for Alpaca and 0.5 for ShareGPT, while gradually increasing the request rate to model 7.
We report individual performance of the heavily loaded model in Figure~\ref{fig:workload}.
At lower request rates, \sysname shows virtually no improvement over vLLM, since the workload is less memory-bound and rarely encounters memory insufficiency. As the request rate increases, however, the improvement becomes significant, exceeding 50\% for both LLaMA and OPT models on Alpaca workloads at request rates of 30 and 20, respectively. Beyond these points, the gains flatten as performance becomes compute-bound.  
For ShareGPT, which is more memory-intensive due to longer output lengths, the improvements continue to grow steadily with higher request rates.  

Across all settings, \sysname achieves 96\% of the performance of \pisysname{}. \pisysname{} has equal access to remote and local memory, making it equivalent to vLLM with a 2$\times$ larger KV cache. This provides an optimal baseline: memory borrowing is maximized in \pisysname{}, where only the heavily loaded GPU consumes memory while all other devices remain free. By contrast, \sysname delivers comparable performance despite having no guarantee of access to remote memory.
This observation raises a key question: why does the presence of lightly loaded models have no measurable effect on the performance gains of their more heavily loaded counterparts? 

\begin{figure}
    \centering
    \includegraphics[width=0.999\linewidth]{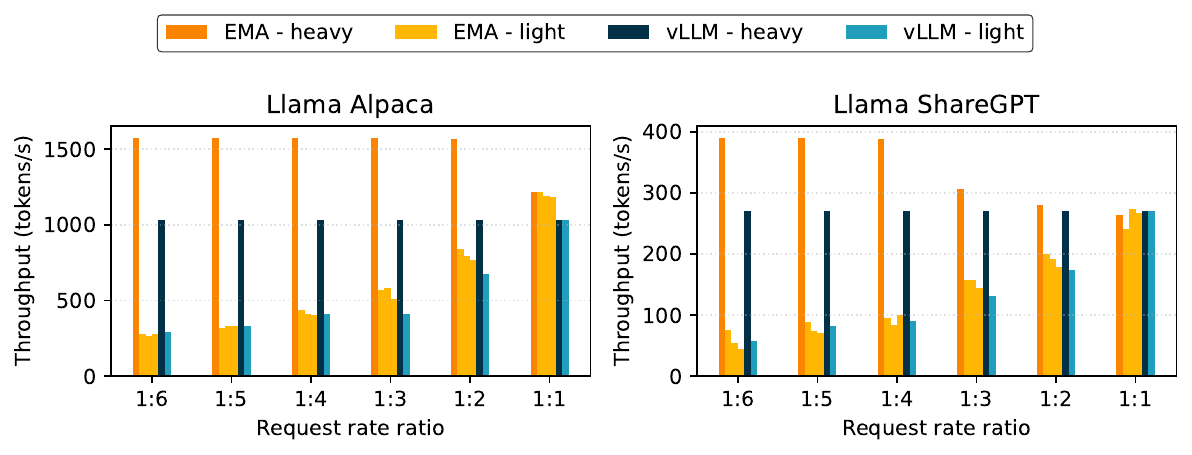}
    \captionsetup{justification=centering}
    \caption{Throughput vs. Request Rate on Lightly-Loaded GPUs}
    \label{fig:tputchange-light}
\end{figure}

We analyze how the improvement changes when the load on lightly loaded GPUs (vLLM and EMA-light) is varied, as shown in Figure~\ref{fig:tputchange-light}. In this setup, the heavily loaded GPU receives a fixed request rate of 30 for Alpaca and 2.5 for ShareGPT (we report only LLaMA results due to space limitations; OPT follows the same trend). Because the request rate of the heavily loaded model does not change, its memory and compute intensity remain constant throughout the workload. Consequently, vLLM-heavy performance remains unchanged, and EMA-heavy improvement depends only on the effectiveness of memory sharing.  

For Alpaca, the load on lightly loaded models has no impact on the heavily loaded GPU, even when the lightly loaded model itself becomes memory-bound (e.g., at ratio 1:2 with request rate 15). The improvement remains above 50\% across ratios from 1:6 to 1:2. A reduction is observed only when the ratio reaches 1:1, where all GPUs operate under the same load, causing the improvement to drop to 20\%.  

For ShareGPT, the heavily loaded GPU is more sensitive to the load on lightly loaded models. Its improvement begins to decrease at a ratio of 1:3. In addition, the lightly loaded models achieve smaller or no improvement.  

These results indicate that memory sharing improves performance on both lightly and heavily loaded GPUs, and the gains are not achieved at the expense of other devices. Sharing is more effective when workloads have moderate memory requirements, as opportunities diminish when all devices operate under high memory pressure. Furthermore, because sharing is performed at request granularity, longer contexts require larger memory chunks, which are difficult to allocate under high memory usage. As a result, when memory utilization is high, sharing is more effective for shorter requests.

\subsection{Memory Utilization and Performance Breakdown}
\label{subsec:eval-mem-util}

\begin{figure}[htbp]
    \centering
    \begin{subfigure}[b]{0.23\textwidth}
        \includegraphics[width=\linewidth]{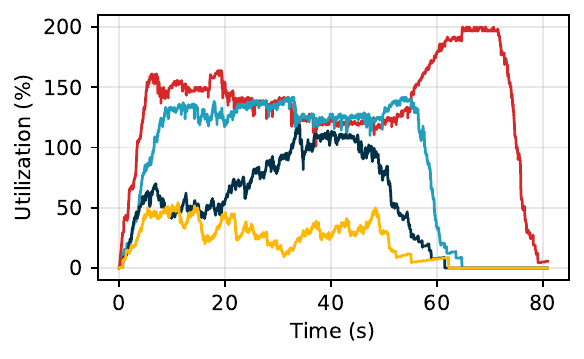}
        \caption{Per GPU Memory Utilization}
        \label{subfig:per-gpu-util}
    \end{subfigure}
    \hfill
    \begin{subfigure}[b]{0.23\textwidth}
        \includegraphics[width=\linewidth]{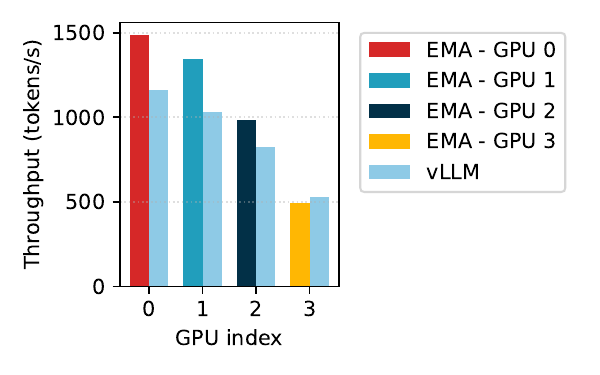}
        \caption{Per GPU Throughput} 
        \label{subfig:per-gpu-tput}
    \end{subfigure}
        \caption{Higher Memory Utilization, Higher Throughput} 
        \label{fig:memutil-tput}
\end{figure}

In this section, we analyze where \sysname's performance gains come from. 
We measure per-GPU memory utilization and throughput on a node with four A100-40GB GPUs, where each GPU hosts a LLaMA-13B model. Alpaca requests are issued to the four GPUs at request rates of 40, 30, 20, and 10 requests/s, respectively. The results are shown in Figure~\ref{fig:memutil-tput}.  

From this experiment, we have three key findings.  
First, borrowing memory has little to no impact on the lender’s performance, which shows the effectiveness of static–dynamic symmetry. Second, we show that the throughput improvement results directly from an increase in effective memory capacity. Among the peers within a node, more heavily loaded GPUs are able to borrow additional memory from underutilized ones. In some cases, a GPU's memory utilization can reach up to 200\%, where 100\% corresponds to its physical HBM capacity and the additional 100\% comes from memory borrowed from its peers. This redistribution leads to more efficient memory allocation across the system: heavily loaded GPUs gain access to more memory, while lightly loaded GPUs are limited to only what they need.
Third, given the dynamic nature of the system, even GPUs that are heavily loaded may experience periods of lower memory utilization. During these moments, other peers are allowed to temporarily use the available memory. As a result, even lightly loaded GPUs can benefit and achieve better performance compared to a baseline where each GPU statically owns and uses only its local memory.
On GPU 0, GPU 1 and GPU 2, \sysname achieves improvements of 1.3$\times$, 1.3$\times$ and 1.2$\times$. GPU 3 remains 0.95$\times$ of the static version performance. The 5\% overhead mostly comes from memory allocation.

\subsection{Latency Comparison}
\label{subsec:latency}

Remote memory sharing is traditionally known to trade off latency for higher throughput. However, \sysname achieves higher throughput without compromising latency by ensuring local--remote symmetry.

\begin{figure}[htbp]
    \centering
    \begin{subfigure}[b]{0.23\textwidth}
        \includegraphics[width=\linewidth]{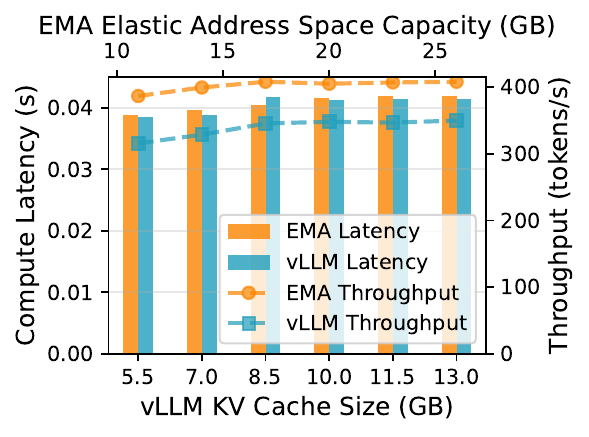}
        \captionsetup{justification=centering}
\caption{LLaMA13B - ShareGPT \\ on 4$\times$A100-40GB}
        \label{fig:llama-latency-expansion}
    \end{subfigure}
    \hfill
    \begin{subfigure}[b]{0.23\textwidth}
        \includegraphics[width=\linewidth]{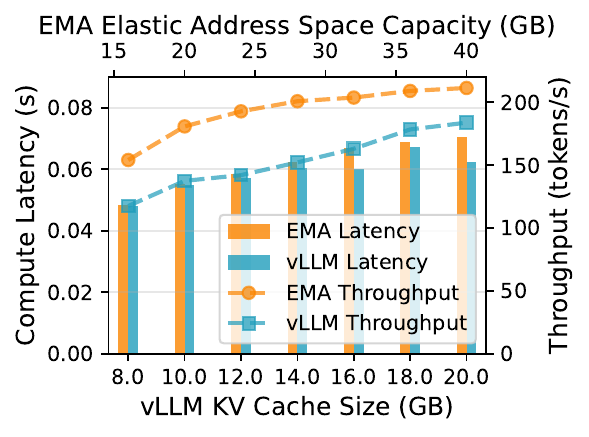}
        \captionsetup{justification=centering}
\caption{OPT30B - ShareGPT \\ on 8$\times$A100-80GB}
        \label{fig:opt-latency-expansion}
    \end{subfigure}
        \caption{Performance Comparison with Larger Cache Size} 
        \label{fig:latency}
\end{figure}

In this section, we compare vLLM and \sysname{} under the same amount of on-device HBM. As shown in Figure~\ref{fig:latency}, \sysname{} provides an Elastic Address Space (EAS) that doubles the available on-device KV-cache capacity. With 4$\times$A100-40GB GPUs running the 13B model, each GPU reaches up to 26\,GB, and with 8$\times$A100-80GB GPUs running the 30B model, each GPU reaches 40\,GB. (The corresponding model sizes are 26.6\,GB and 58.8\,GB.)

We find that \sysname{} maintains local--remote symmetry by sustaining nearly the same compute latency while varying the expanded memory size. With large expansions (e.g., 20\,GB on A100-80GB), \sysname{} shows a 10\% increase in latency. This is primarily due to larger batch sizes enabled by the expanded capacity, which increase per-iteration compute time, and DMA synchronization and coordination overhead between the memory slice processes. Even so, the added capacity to EAS still improves overall throughput.

\subsection{Local--Remote Symmetry}
\label{subsec:eval-lr}

In this section, we evaluate the importance of local–remote symmetry in memory sharing and quantify the performance improvement this property provides.

We implement a variant of \sysname that forgoes local–remote symmetry. Rather than prefetching remote data ahead of use, it performs on-demand swapping of an entire KV-cache layer only when accessed. 
Compared to HUVM~\cite{miao2022humv}, this variant offers a stronger design under a GPU-only memory setting. While HUVM services page faults incrementally at the granularity of individual pages, our variant transfers the full layer in a single operation, allowing one large memcpy to eliminate the substantial per-page overhead of on-demand paging.

\begin{figure}[htbp]
    \centering
    \begin{subfigure}[b]{0.23\textwidth}
        \includegraphics[width=\linewidth]{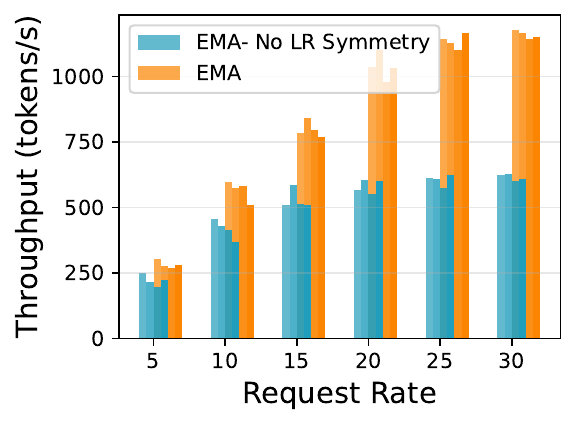}
        \captionsetup{justification=centering}
        \caption{LLaMA-13B - Alpaca \\Throughput}
        \label{fig:lr-tput}
    \end{subfigure}
    \hfill
    \begin{subfigure}[b]{0.23\textwidth}
        \includegraphics[width=\linewidth]{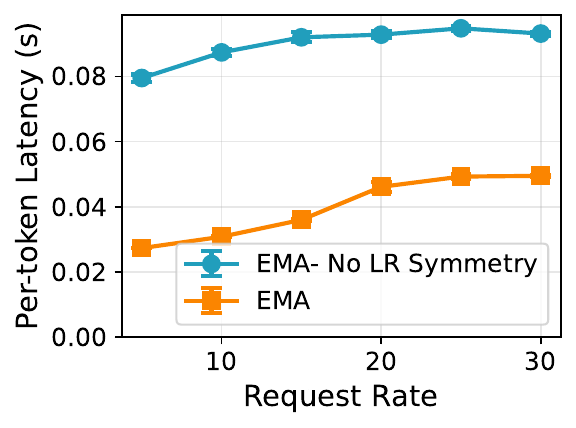}
        \captionsetup{justification=centering}
        \caption{LLaMA-13B - Alpaca \\ Per Token Compute Latency} 
        \label{fig:lr-latency}
    \end{subfigure}
        \caption{Local–Remote Symmetry} 
        \label{fig:lr-symmetry}
\end{figure}

From Figure~\ref{fig:lr-symmetry}, we find that local-remote symmetry is important in both throughput and latency, as computation must stall whenever accessed data resides on remote memory and must be swapped in on the critical path. 

\subsection{Static--Dynamic Symmetry}
\label{subsec:eval-sd}

\sysname ensures that each GPU, when underutilized, contributes its free memory to memory slices for sharing. However, when it requires any of the contributed memory back, it must be able to logically reclaim it immediately.
This subsection evaluates the performance impact of enforcing the static–dynamic symmetry guarantee.

We compare the performance of \sysname with and without this guarantee. Specifically, we report the individual GPU throughput, the number of preemptions caused by out-of-memory conditions, and the number of preemptions triggered by memory reclamation.
We find that with an effective reclamation strategy, \sysname{} with static–dynamic symmetry achieves aggregate throughput within –5\% to +10\% relative to the version without it, while additionally providing the symmetry guarantee.

\begin{figure}[htbp]
    \centering
    \begin{subfigure}[b]{0.23\textwidth}
        \includegraphics[width=\linewidth]{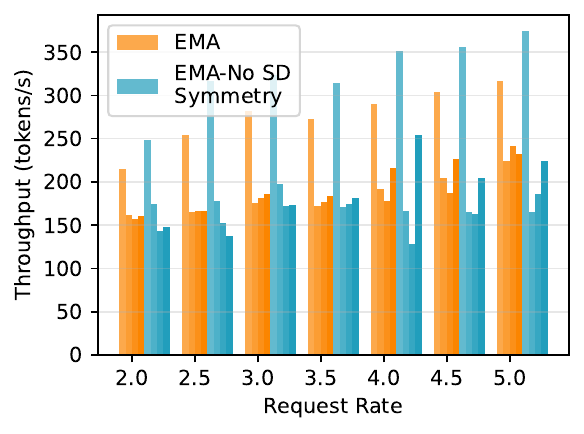}
        \captionsetup{justification=centering}
        \caption{LLaMA-13B - ShareGPT \\ Throughput}
        \label{fig:sd-tput}
    \end{subfigure}
    \hfill
    \begin{subfigure}[b]{0.23\textwidth}
        \includegraphics[width=\linewidth]{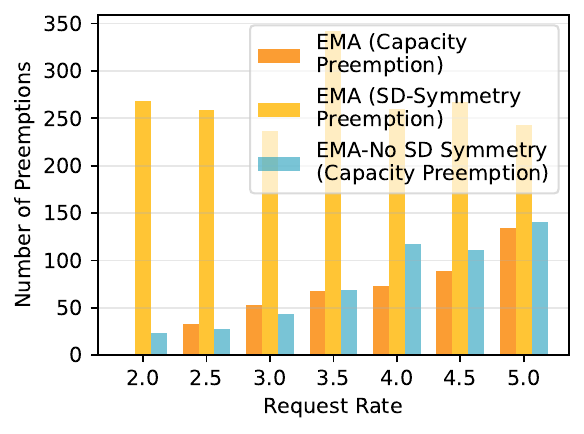}
        \captionsetup{justification=centering}
        \caption{LLaMA-13B - ShareGPT  \\ Number of Preemptions}
        \label{fig:sd-preemptions}
    \end{subfigure}
        \caption{Static-Dynamic Symmetry}
        \label{fig:static-dynamic-eval}
\end{figure}

\subsection{Maximizing Performance per Slice}
\label{subsec:eval-slice-setup}

\begin{figure}[htbp]
    \centering
    \begin{subfigure}[b]{0.23\textwidth}
        \includegraphics[width=\linewidth]{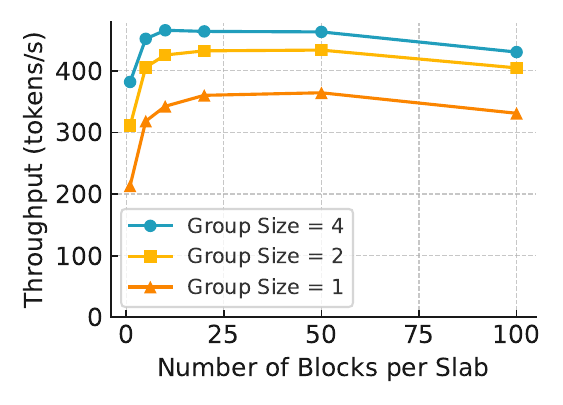}
        \caption{OPT-13B}
        \label{fig:intraslice13b}
    \end{subfigure}
    \hfill
    \begin{subfigure}[b]{0.23\textwidth}
        \includegraphics[width=\linewidth]{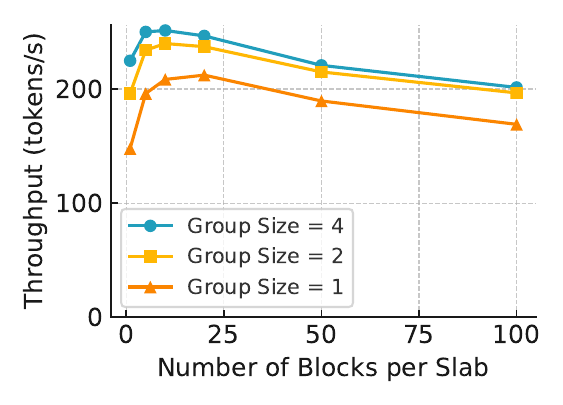}
        \caption{OPT-30B} 
        \label{fig:intraslice30b}
    \end{subfigure}
        \caption{Throughput under Different Group and Slab Size} 
        \label{fig:intraslice}
\end{figure}

\sysname relies on the abstraction of memory slices. Before the system becomes compute-bound, increasing the combined capacity of memory slices always leads to higher performance.
In this section, we study how to configure a memory slice to achieve higher performance with the same capacity.
Figure~\ref{fig:intraslice} shows the throughput of OPT models evaluated on ShareGPT.

We fix the layer group size, and vary the slab size from 1 to 100.
We find that when the slab size is small, \sysname frequently stalls at each layer boundary, resulting in higher per-token latency and reduced throughput.
On the other hand, with a large slab size, slab utilization and the opportunity for memory sharing are significantly compromised. This reduces the effective memory available to each model and ultimately lowers throughput.

We also study the impact of layer group size by varying it from 1 to 4.
A smaller group size leads to performance degradation for the same reason as smaller slab size: the smaller granularity of data transfer results in lower effective bandwidth and increased waiting time.
However, a group size that is too large leads to memory inefficiency, as it requires larger swap-in/out buffers. The buffer size must accommodate one entire group of layers in flight, which increases memory overhead and reduces available space for actual usage.
Therefore, we find that a group size of 4 and a slab size of 10 usually yield consistently good performance across workloads and models, and we set these as the default values.

%% file: related.tex
\section{Related Work}
\paragraph{General Model Serving Systems.}
Prior work aims to improve throughput, latency, and elastic scaling for inference pipelines~\cite{li2023alpaserve,crankshaw2017clipper,olston2017tensorflow,shen2019nexus,crankshaw2020inferline,zhang2023shepherd,yu2022orca}.
Transformer stacks such as FasterTransformer, LightSeq, TurboTransformers, DeepSpeed-Inference, and Accelerate optimize kernels, scheduling, and batching to increase single-node efficiency~\cite{nvidia_fastertransformer,wang2020lightseq,fang2021turbotransformers,aminabadi2022deepspeed,huggingface_accelerate}.
These systems treat GPU memory as a fixed resource, focusing on \emph{how} it is used rather than \emph{how much} is available.
In practice, capacity pressure, especially from LLM KV caches, still forces queuing, preemption, or model downsizing, motivating mechanisms that expand usable memory rather than only improve scheduling.

\paragraph{Recomputation and Offloading/Swapping to CPU.}
To stretch capacity without new hardware, many approaches trade extra compute or I/O for memory headroom.
Recomputation and checkpointing reduce peak activation footprint during training~\cite{chen2016training,herrmann2019optimal,korthikanti2023reducing,jain2020checkmate,dao2022flashattention,dao2023flashattention},
while swapping pipelines move tensors between GPU and host or NVMe~\cite{abhyankarinfercept, sheng2023flexgen, kwon2023efficient, ren2021zero,huang2020swapadvisor,
rhu2016vdnn,wang2018superneurons,
peng2020capuchin,yang2024protrain}.
All of these are fundamentally reactive: data is brought back only when an access occurs.
This makes expanded capacity helpful but not transparent—applications observe a clear performance delta between local HBM and slower tiers.

\paragraph{Swapping/Sharing Between GPUs.}
A complementary direction borrows memory from peer GPUs.
HUVM demonstrates that idle HBM on neighboring devices can be harvested over high-bandwidth interconnects to expand effective capacity~\cite{miao2022humv}, and object-aware managers further explore multi-GPU placement and migration policies~\cite{jiang2025oasis}.
AQUA~\cite{vijayakumar2025aqua} enables preemptive prompt scheduling by offloading KV cache to peer GPUs over NVLink instead of host DRAM, using a profiler to classify GPUs as memory producers or consumers.
Harvest~\cite{gopal2026harvest} treats unused peer HBM as an opportunistic cache tier for MoE expert weights and KV cache entries, retrieving them over GPU-to-GPU interconnects on demand.
However, these designs still move data reactively, whether triggered by scheduling-driven page-ins or cache misses, so remote HBM operates as a faster-than-CPU but still observably slower tier rather than a truly uniform extension of local memory.
By contrast, \sysname treats peer memory as a near tier that can be made \emph{performance-transparent} during compute by proactively prefetching data ahead of use, and additionally guarantees that lenders can reclaim shared memory on demand without performance loss.

%% file: conclusion.tex
\section{Conclusion}
\label{sec:conclusion}

In conclusion, we present the objectives and principles of cross-GPU memory sharing and introduce \sysname{}. By exploiting predictable memory access patterns, the high bandwidth of modern interconnects, and real-time information, \sysname{} optimizes memory utilization in unbalanced and highly dynamic environments. It allows each user to transparently access peers' memory with performance indistinguishable from local memory, thereby achieving higher throughput without sacrificing latency.

%% file: appendix.tex
\section*{Appendix}
\appendix
\section{Proof of Capacity, Latency and Bandwidth Relationships}
\label{sec:proof}

To prove that the capacity requirement in a memory slice must be satisfied to enable local--remote symmetry in memory sharing, we begin with the following four inequalities:

\begin{align}
B_{j \rightarrow i}^i \cdot L_i &> m_{ji} \tag{1} \\
B_{i \rightarrow j}^j \cdot L_j &> m_{ij} \tag{2} \\
B_{i \rightarrow j}^i \cdot L_i &> m_{ji} \tag{3} \\
B_{j \rightarrow i}^j \cdot L_j &> m_{ij} \tag{4}
\end{align}

These indicate that each GPU must be able to transmit and receive the same amount of data in both directions.

We also have bandwidth decomposition equations:

\begin{align}
B_{j \rightarrow i}^i + B_{j \rightarrow i}^j &= B_{j \rightarrow i} \tag{5} \\
B_{i \rightarrow j}^i + B_{i \rightarrow j}^j &= B_{i \rightarrow j} \tag{6}
\end{align}

Let us define:

\[
\alpha = \frac{B_{j \rightarrow i}^i}{B_{j \rightarrow i}}, \quad
\beta = \frac{B_{i \rightarrow j}^j}{B_{i \rightarrow j}}
\]

Then:

\[
B_{j \rightarrow i}^i = \alpha B_{j \rightarrow i}, \quad
B_{j \rightarrow i}^j = (1 - \alpha) B_{j \rightarrow i}
\]
\[
B_{i \rightarrow j}^j = \beta B_{i \rightarrow j}, \quad
B_{i \rightarrow j}^i = (1 - \beta) B_{i \rightarrow j}
\]

Substituting into inequalities (1)--(4), we get:

\begin{align*}
\alpha B_{j \rightarrow i} \cdot L_i &> m_{ji} \\
(1 - \alpha) B_{j \rightarrow i} \cdot L_j &> m_{ij} \\
(1 - \beta) B_{i \rightarrow j} \cdot L_i &> m_{ji} \\
\beta B_{i \rightarrow j} \cdot L_j &> m_{ij}
\end{align*}

From these, we obtain constraints:

\[
\frac{m_{ji}}{B_{j \rightarrow i} L_i} < \alpha < 1 - \frac{m_{ij}}{B_{j \rightarrow i} L_j}
\quad \Rightarrow \quad
\frac{m_{ji}}{L_i} + \frac{m_{ij}}{L_j} < B_{j \rightarrow i} \tag{A'}
\]
\[
\frac{m_{ij}}{B_{i \rightarrow j} L_j} < \beta < 1 - \frac{m_{ji}}{B_{i \rightarrow j} L_i}
\quad \Rightarrow \quad
\frac{m_{ji}}{L_i} + \frac{m_{ij}}{L_j} < B_{i \rightarrow j} \tag{B'}
\]

Therefore, a necessary and sufficient condition for all inequalities (1)--(4) to hold is:

\[
\boxed{
\frac{m_{ji}}{L_i} + \frac{m_{ij}}{L_j} < \min(B_{j \rightarrow i}, B_{i \rightarrow j})
}
\]

\bigskip

\noindent\textbf{Tightest Case.}  
When the above inequality becomes equality, we achieve the tightest case:

\[
\frac{m_{ji}}{L_i} + \frac{m_{ij}}{L_j} = B_{j \rightarrow i} = B_{i \rightarrow j} = B
\]

\section{Proof of Per GPU Contribution to Memory Slice}
\label{proof:per-gpu-contribution}

\paragraph{Proposition.}
Let $m_{ji},m_{ij}\geq 0$ and denote $S=m_{ji}+m_{ij}$. Suppose that for any $m_v\in[0,S]$,
\[
\frac{m_{ji}-m_v}{L_{ir}}+\frac{m_{ij}+m_v}{L_{jr}}\leq B, 
\qquad
\frac{m_{ji}+m_v}{L_{ir}}+\frac{m_{ij}-m_v}{L_{jr}}\leq B,
\]
where $L_{ir}\leq L_i$ and $L_{jr}\leq L_j$ are individual application latency windows. Then
\[
S_{\max}(L_{ir},L_{jr}) = B\cdot \min(L_{ir},L_{jr}).
\]
Moreover,
\[
\max_{\substack{L_{ir}\leq L_i \\ L_{jr}\leq L_j}} S_{\max}(L_{ir},L_{jr})
= B \cdot \min(L_i,L_j),
\]
achieved when
\[
L_{ir}=L_{jr}=\min(L_i,L_j).
\]

\paragraph{Proof.}
For any fixed $(m_{ji},m_{ij})$ with total $S=m_{ji}+m_{ij}$, each inequality can be written as
\[
\frac{m_{ji}}{L_{ir}}+\frac{m_{ij}}{L_{jr}} 
\;\pm\; m_v\!\left(\frac{1}{L_{ir}}-\frac{1}{L_{jr}}\right).
\]
Requiring both inequalities for all $m_v\in[0,S]$ is equivalent to
\begin{equation}
\frac{m_{ji}}{L_{ir}}+\frac{m_{ij}}{L_{jr}} 
+ S \left|\tfrac{1}{L_{ir}}-\tfrac{1}{L_{jr}}\right|
\;\leq\; B.
\label{eq:robust}
\end{equation}

For a given total $S$, the term $\tfrac{m_{ji}}{L_{ir}}+\tfrac{m_{ij}}{L_{jr}}$ is minimized by assigning all of $S$ to the side with the larger reported rate. Without loss of generality, assume $L_{ir}\geq L_{jr}$. Then the minimum is $\tfrac{S}{L_{ir}}$, and since $\bigl|\tfrac{1}{L_{ir}}-\tfrac{1}{L_{jr}}\bigr| = \tfrac{1}{L_{jr}}-\tfrac{1}{L_{ir}}$, the left-hand side of \eqref{eq:robust} simplifies to
\[
\frac{S}{L_{ir}} + S\Bigl(\frac{1}{L_{jr}}-\frac{1}{L_{ir}}\Bigr) = \frac{S}{L_{jr}}.
\]
Thus feasibility requires $S/L_{jr}\leq B$, i.e.,
\[
S\leq B\,L_{jr}=B\,\min(L_{ir},L_{jr}).
\]
This bound is tight: with $L_{ir}\geq L_{jr}$, the allocation $(m_{ji},m_{ij})=(S,0)$ achieves equality when $S=B L_{jr}$. Therefore,
\[
S_{\max}(L_{ir},L_{jr})=B\,\min(L_{ir},L_{jr}).
\]

Finally, since $L_{ir}\leq L_i$ and $L_{jr}\leq L_j$, maximizing over all admissible reported rates gives
\[
\max_{\substack{L_{ir}\leq L_i \\ L_{jr}\leq L_j}} S_{\max}(L_{ir},L_{jr})
= B\,\min(L_i,L_j),
\]
which is attained by setting $L = L_{ir}=L_{jr}=\min(L_i,L_j)$, then:

\[
\boxed{
m_{ji} + m_{ij} = B \cdot L
}
\]

\section{Proof of Scheduling Algorithm}
\label{sec:proof-scheduling}
\paragraph{Informal description.}
We formulate allocation on memory slices as an online scheduling problem.
Each slice has normalized capacity~1.

\medskip
\paragraph{Setting.}
Let $m$ be the number of GPUs, $V=\{1,\dots,m\}$.
For every unordered pair $\{i,j\}\subset V$ with $i\neq j$, there is a shared memory slice of normalized capacity~$1$ that both GPUs can allocate from.
The utilization of slice $\{i,j\}$ at time $t$ equals its current allocation (since capacity is $1$):
\[
u_{\{i,j\}}(t)\;=\; \text{alloc}_{\{i,j\}}(t)\;\in[0,1].
\]

\paragraph{Jobs and demand.}
Requests (``jobs'') arrive online; each job $r$ is owned by a GPU $i_r\in V$, has an arrival time $a_r\in\mathbb{Z}_{\ge 0}$, and a time-varying (possibly non-monotone) demand function
\[
d_r:\mathbb{Z}_{\ge 0}\to \mathbb{Z}_{\ge 0},\qquad d_r(t)=\text{units held by }r\text{ at time }t.
\]
At each tick $t$, the allocator observes $\Delta d_r(t)=d_r(t)-d_r(t-1)$.
If $\Delta d_r(t)>0$, that many new units must be placed on slices $\{i_r,j\}$ with $j\neq i_r$;
if $\Delta d_r(t)<0$, that many units are released.
A job departs once $d_r(t)=0$ for all subsequent $t$.

\paragraph{Online allocation objective.}
At each time $t$, the allocator must irrevocably assign every positive increment to a slice incident to the owner GPU, without knowledge of future demands.
We track:
\[
\textsf{Cong}(A)=\sup_{t\ge 0}\,\max_{\{i,j\}} u^{A}_{\{i,j\}}(t)
\quad\text{(peak utilization under allocator $A$)},
\]
and the number of slice-saturation events
\[
S^{\max}(A)=\#\bigl\{(t,\{i,j\})\ \big|\ u_{\{i,j\}}^{A}(t^-)<1,\ u_{\{i,j\}}^{A}(t)=1\bigr\}.
\]
The competitive ratio compares an online allocator $A$ to an offline optimal $\textsf{OPT}$ that knows all arrivals, releases, and demands in advance:
\[
\rho(A)=\sup_{\text{demand sequences}}\frac{\textsf{Cong}(A)}{\textsf{Cong}(\textsf{OPT})}.
\]

\medskip
\paragraph{Greedy–Least-Loaded (GLL).}
When GPU~$i$ receives a unit, it scans its $m\!-\!1$ incident slices $\{i,j\}$ and chooses the one with minimum utilization $u_{\{i,j\}}(t)$.  
This requires only per-slice counters adjacent to $i$ and $O(m)$ work per decision.

Azar \& Epstein~\cite{azai1997line, azar2005convex} prove
\begin{align*}
  \max_{t,\{i,j\}} u^{\text{GLL}}_{\{i,j\}}(t) &\;\le\; \tfrac{m+3}{2}\,
  \max_{t,\{i,j\}} u^{\textsf{OPT}}_{\{i,j\}}(t) \\[6pt]
  \Longrightarrow\quad
  S^{\max}\!\bigl(\text{GLL}\bigr) &\;\le\; \tfrac{m+3}{2}\,
  S^{\max}(\textsf{OPT}).
\end{align*}

Thus GLL may cause \emph{up to $\Theta(m)$ more slices to saturate} than an omniscient allocator, yet it needs only local state and $O(m)$ work per decision.

\paragraph{Interpretation.}
The worst case arises from adversarial demand sequences that force myopic early placements to later coalesce on the same slices, causing avoidable saturations that an omniscient allocator could have steered around.
Despite this $\Theta(m)$ worst-case gap, GLL is appealing in practice because it is decentralized, needs only local state, and incurs low decision overhead.